\RequirePackage[2024-06-01]{latexrelease}
\documentclass[aip,cha,reprint,amsmath,amssymb,nofootinbib,floatfix]{revtex4-2}

\usepackage{graphicx,bm,booktabs,amsthm}
\usepackage[most]{tcolorbox}
\usepackage{array}
\usepackage{needspace}
\usepackage{colortbl}
\definecolor{lbblue}{RGB}{225,236,250}
\definecolor{ubred}{RGB}{252,226,226}
\newtcolorbox{ideabox}[2][]{breakable,enhanced,colback=blue!4,colframe=blue!45!black,boxrule=0.5pt,arc=1.5pt,
  left=5pt,right=5pt,top=4pt,bottom=4pt,fontupper=\small,title={#2},fonttitle=\small\bfseries,
  coltitle=blue!45!black,colbacktitle=blue!4,titlerule=0pt,toptitle=1pt,bottomtitle=0pt,#1}
\newtheorem{theorem}{Theorem}
\newtheorem{lemma}{Lemma}
\newtheorem{proposition}{Proposition}

\newtheorem{definition}{Definition}

\newcommand{\E}{\mathcal{E}}
\newcommand{\B}{\mathcal{B}}
\newcommand{\dd}{\mathrm{d}}
\newcommand{\mus}{\mu_*}
\newtcolorbox{conjbox}[1][]{enhanced,unbreakable,colback=yellow!30,colframe=red,boxrule=0.6pt,
  sharpish corners,left=5pt,right=5pt,top=3pt,bottom=3pt,fontupper=\small,title={Conjecture},
  fonttitle=\small\bfseries,center title,#1}

\begin{document}

\title{Below 3/4: the critical connectivity of Kuramoto networks is at most 0.717}

\author{Kevin P. O'Keeffe}
\email{kevin.p.okeeffe@gmail.com}
\affiliation{Starling Research Institute, Seattle, WA 98112, USA}

\date{\today}


\begin{abstract}
How densely must a network of identical Kuramoto oscillators be wired before synchrony is its only
stable state? Kassabov, Strogatz and Townsend showed that connectivity $3/4$ suffices, and Lepsveridze and
Zhang that the true threshold $\mu_c$ lies strictly below $3/4$, without saying by how much. We prove the
first explicit bound below $3/4$: $\mu_c\le0.717$. With our recent lower bound, this pins $\mu_c$ to
$[0.6929,0.717]$. The obstacle is a family of ``square states,'' with every oscillator at a corner of one
square, where linear stability analysis can be blind. We work inward: exact rational certificates rule out
states far from them, an explicit trial rotation rules out states near them, and a cubic calculation
settles the square states themselves.
\end{abstract}

\maketitle

\begin{quotation}
\noindent\textbf{Networks of identical oscillators that are densely wired tend to synchronize. How dense
is dense enough? For the Kuramoto model there is a critical connectivity $\mu_c$: once every oscillator
is coupled to more than a fraction $\mu_c$ of the others, sync is the only stable equilibrium, meaning the
only local minimum of the energy. Kassabov, Strogatz and Townsend proved that $\mu_c$ is at most $3/4$, and Lepsveridze and Zhang
recently showed that it is strictly smaller, though by an unknown amount. The trouble lies at a special state: four equal clusters a quarter turn apart, balanced on a
razor's edge. We prove that $\mu_c\le0.717$, working inward. States far from the razor's edge are ruled out by
computer-checked certificates in exact arithmetic. The hard part is the states that hug it: for each one
we build a trial rotation that lowers the energy. States on the razor's edge itself are ruled out by a
short cubic calculation.}
\end{quotation}

\begin{table*}
\caption{Bounds on the critical connectivity $\mu_c$. Lower bounds (left) are networks with stable
nonsynchronous states; upper bounds (right) are theorems that sync must win. The upper bounds of
Ref.~\onlinecite{LZ2026} and this work concern local minima; Ref.~\onlinecite{KST2021} also proves
convergence to sync from almost every initial condition. WSG networks are the nearest-neighbor rings of
Wiley, Strogatz and Girvan~\cite{WSG2006}.}
\label{tab:history}
\small
\centering
\setlength{\tabcolsep}{4pt}
\begin{tabular}{@{\extracolsep{0pt}}>{\columncolor{lbblue}[\tabcolsep][1.5\tabcolsep]}l>{\columncolor{lbblue}[1.5\tabcolsep][1.5\tabcolsep]}l>{\columncolor{lbblue}[1.5\tabcolsep][1.5\tabcolsep]\raggedright\arraybackslash}p{0.22\textwidth}>{\columncolor{lbblue}[1.5\tabcolsep][\tabcolsep]}l|>{\columncolor{ubred}[\tabcolsep][1.5\tabcolsep]}l>{\columncolor{ubred}[1.5\tabcolsep][1.5\tabcolsep]}l>{\columncolor{ubred}[1.5\tabcolsep][1.5\tabcolsep]\raggedright\arraybackslash}p{0.22\textwidth}>{\columncolor{ubred}[1.5\tabcolsep][\tabcolsep]}l}
\hline\hline
\multicolumn{4}{>{\columncolor{lbblue}}c|}{lower bounds} & \multicolumn{4}{>{\columncolor{ubred}}c}{upper bounds}\\
year & bound & method & ref. & year & bound & method & ref.\\
\hline
2006 & $\mu_c\ge0.6809$ & twisted states on rings & \onlinecite{Taylor2012,WSG2006} &
2012 & $\mu_c\le0.9395$ & necessary condition for stable fixed points & \onlinecite{Taylor2012}\\
2015 & $\mu_c\ge0.6818$ & lexicographic product of WSG and complete graphs & \onlinecite{CanaleMonzon2015} &
2019 & $\mu_c\le0.7929$ & landscape analysis & \onlinecite{Ling2019}\\
2020 & $\mu_c\ge0.6828$ & twinning on WSG networks & \onlinecite{TSS2020} &
2020 & $\mu_c\le0.7889$ & refined estimates & \onlinecite{LuSteinerberger2020}\\
2021 & $\mu_c\ge0.6838$ & optimal circulants & \onlinecite{Yoneda2021} &
2021 & $\mu_c\le3/4$ & order parameters $\rho_1,\rho_2$ & \onlinecite{KST2021}\\
2022 & $\mu_c\ge11/16$ & eight-cluster construction & \onlinecite{Canale2022} &
2026 & $\mu_c<3/4$ & nonlinear stability, no estimate & \onlinecite{LZ2026}\\
2026 & $\mu_c\ge0.6929077$ & split four-cluster states & \onlinecite{OKeeffeLB} &
this work & $\mu_c\le0.717$ & trial rotation + certificates & \\
\hline\hline
\end{tabular}
\end{table*}

%==========================================================================================
\section{Introduction}\label{sec:intro}

Networks of identical, attractively coupled oscillators tend to synchronize. If every oscillator is
coupled to enough of the others, sync is the only stable state~\cite{Taylor2012}. The question is how
many is ``enough''.

Consider the homogeneous Kuramoto model on a simple graph with adjacency matrix $A$,
\begin{equation}
\dot\theta_j=\sum_{k=1}^{n}A_{jk}\sin(\theta_k-\theta_j),\qquad j=1,\dots,n.
\label{eq:kuramoto}
\end{equation}
It is a gradient flow for the potential, or energy,
\begin{equation}
V(\theta)=\sum_{j<k}A_{jk}\bigl[1-\cos(\theta_j-\theta_k)\bigr],
\label{eq:potential}
\end{equation}
and we study its local minima. Throughout this paper a \textit{stable state} means a local minimum of $V$.

Following Refs.~\onlinecite{Taylor2012,KST2021}, we measure how densely a graph is wired by its
\textit{connectivity} $\mu=\delta/(n-1)$ (for $n\ge2$), where $\delta$ is the minimum degree: every oscillator is
coupled to at least a fraction $\mu$ of the others. The critical connectivity $\mu_c$ is the infimum of
the numbers $\mu$ such that every graph with connectivity at least $\mu$ has no stable state other than
sync. This is the local-minimum version of $\mu_c$, as in Refs.~\onlinecite{LZ2026,OKeeffeLB}. A
stronger, ``dynamical'' version asks for convergence to sync from almost every initial
condition~\cite{KST2021}. Our bound, like the strict bound of Ref.~\onlinecite{LZ2026}, is for the
local-minimum version; for the dynamical version, $3/4$ remains the best upper bound. (Our results do not give it, since some equilibria fail only at third order.)

\begin{figure}[t]
\centering
\includegraphics[width=0.68\columnwidth]{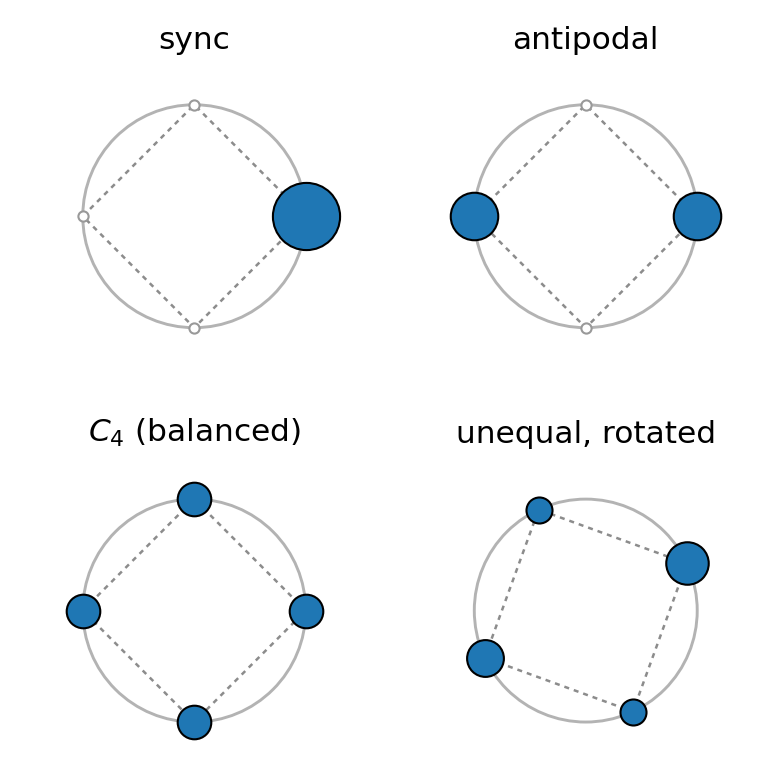}
\caption{Four square states. Every oscillator sits at a corner of one square inscribed in the circle
(dashed); dots are clusters, sized by mass, and open circles are empty corners. The square may be rotated,
and any corners may be occupied, with any masses. Linear stability cannot rule these states out: when no
links join opposite corners, every link is a quarter turn long or lies inside a cluster, so the energy is
flat to second order.}
\label{fig:square}
\end{figure}

Table~\ref{tab:history} collects what is known. Upper bounds are proofs that sync must win, and the best
was $3/4$, by Kassabov, Strogatz and Townsend~\cite{KST2021}. They also found what stands in the way of
going lower: networks with connectivity near $3/4$ that, in their words, ``lie on the so-called razor's edge of
stability,'' with four equal clusters a quarter turn apart (Fig.~\ref{fig:square}). The energy of this state is flat to second order when its clusters are moved
rigidly, so linear stability analysis cannot decide whether it is stable. Lepsveridze and Zhang recently got past the
razor's edge with a nonlinear argument and proved that $\mu_c<3/4$ strictly~\cite{LZ2026}, but their
argument does not say by how much. Our argument near the four-cluster state makes their strategy quantitative.

Lower bounds go the other way: they are networks with stable nonsynchronous states. In a companion paper we
constructed such networks with connectivity arbitrarily close to
$\mu_0=0.6929077\ldots$~\cite{OKeeffeLB}. Their phases sit close to the razor's-edge state: four clusters a
quarter turn apart, three of them split into two slightly separated halves.
So both bounds meet the same special state.

\begin{theorem}[Main result]\label{thm:main}
Let $G$ be a simple graph on $n$ vertices with minimum degree $\delta$, and suppose
$(\delta+1)/n\ge717/1000$. Then every local minimum of the energy $V$ of Eq.~\eqref{eq:potential} on $G$ is synchronized. In particular, this
holds whenever the connectivity $\mu=\delta/(n-1)$ is at least $717/1000$, so $\mu_c\le717/1000$.
\end{theorem}
The $+1$ counts each oscillator's harmless link to itself (Sec.~\ref{sec:setting}), and
$(\delta+1)/n\ge\delta/(n-1)$.
Together with Ref.~\onlinecite{OKeeffeLB},
\begin{equation}
0.6929077\ldots\le\mu_c\le0.717.
\end{equation}
This is the first explicit upper bound below $3/4$; with Ref.~\onlinecite{OKeeffeLB} it narrows the window
for $\mu_c$ from $[11/16,3/4)$ to $[0.6929,0.717]$.

To prove the theorem we must show that every nonsynchronous equilibrium of a dense network is unstable:
its Hessian has a negative eigenvalue or, where the Hessian is degenerate, the energy still drops at
higher order. One fact organizes the proof: \emph{linear stability is neutral at the ``square states,''} in
which every oscillator sits at a corner of one square inscribed in the circle (Fig.~\ref{fig:square}).
On these states, when opposite corners are unlinked, the clusters can be moved rigidly at no cost to second order, so the Hessian has extra zero eigenvalues. Our idea is to sort equilibria by their distance from the square states, measured by $D_4=1-|\langle e^{4i\theta}\rangle|^2$, which vanishes exactly on them (quadrupling the phases sends a square's corners to one point), into three regimes, each with its own technique, working inward.

\emph{Far} from the square states (large $D_4$), we rule out stability altogether using \textit{certificates}. First, we summarize each state by
averages of its phases and links. Then equilibrium, stability and the degree bound become a system of
linear and matrix inequalities, which a convex optimization solver shows has no solution by producing
the certificate: explicit weights that combine the inequalities into a contradiction, checked in exact
arithmetic. Certificates cannot reach the square states, though. At $D_4=0$ there is nothing to
contradict, since $C_4$ satisfies every second-order condition, and just next to it the contradiction
becomes vanishingly weak.

\emph{Near} the square states (small $D_4$), the zero eigenvalues are lifted, but only slightly, so any negative eigenvalue is tiny. We find it with the variational principle. As with a
trial wavefunction in quantum mechanics, one explicit rule builds, for \emph{every} candidate state in the near regime, a perturbation, a trial rotation, that lowers its energy at second order, so each has a negative eigenvalue.
The energy drop is proportional to $D_4$, so it vanishes on the square states.

\emph{On} the square states ($D_4=0$), a cubic term of definite sign decides stability.

%==========================================================================================
\section{From networks to macrostates}\label{sec:setting}

The plan of the Introduction needs three things, which this section supplies. First, a way to remove the
graph altogether, describing a state on any graph, of any size, by one finite object: its
macrostate. Second, the conditions that object must satisfy if the state is stable. Third, the
distance $D_4$ from the square states, which sorts macrostates into the far, near, and on regimes.

\begin{figure}[t]
\centering
\includegraphics[width=\columnwidth]{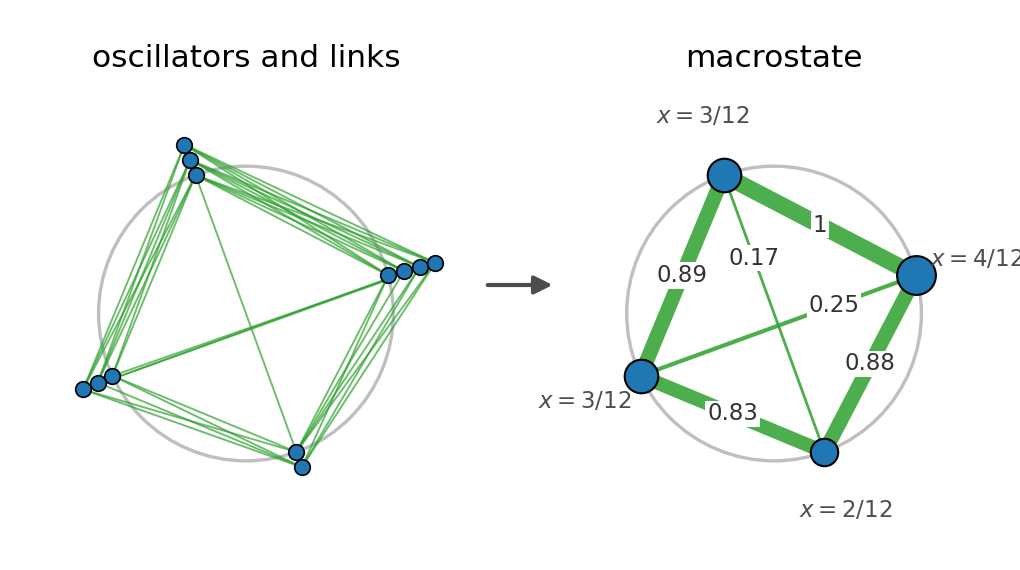}
\caption{From a network state to its macrostate (schematic). Left: twelve oscillators on the unit circle,
drawn as points at their phases, with a chord for every edge of the network; oscillators with the same phase are stacked
outward so that they can be seen. Right: each stack becomes one dot of mass $x_a$, and each bundle of
chords becomes one line whose label and width give the link density $P_{ab}$, the fraction of possible
links that are present. Links inside a stack are not drawn.}
\label{fig:coarse}
\end{figure}

\begin{figure*}[t]
\includegraphics[width=0.92\textwidth]{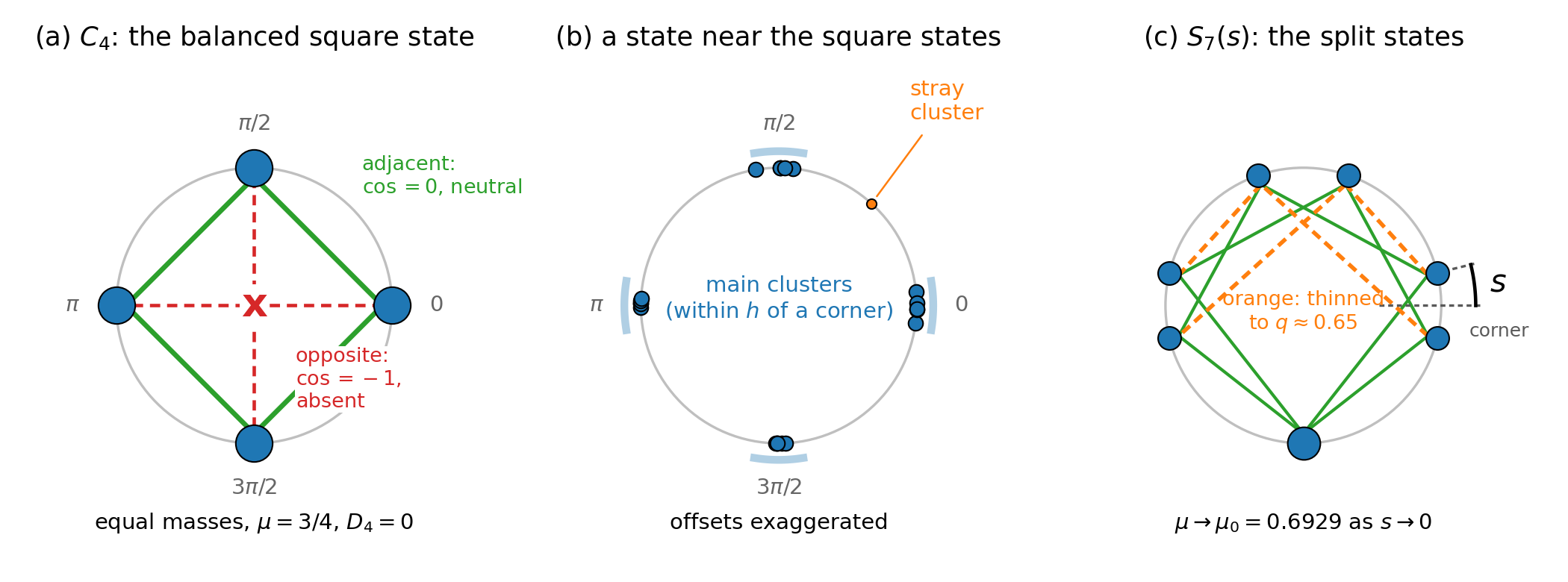}
\caption{Three running examples (schematic; sizes and offsets exaggerated). (a) $C_4$: equal clusters at
the four corners, adjacent clusters fully linked (green), opposite clusters unlinked (red); its degree is
$3/4$. (b) A state near the square states: main clusters in thin groups at the corners, plus a stray
cluster. (c) The split states $S_7(s)$ of Ref.~\onlinecite{OKeeffeLB}, drawn with exaggerated offsets and their limiting link pattern: three neighboring clusters split into halves about $s$ either side of their corner (the middle one about $1.4s$), with the four crossed links between halves thinned to density $q\approx0.65$ (orange). Their degree
approaches $\mu_0=0.6929\ldots$ as $s\to0$.}
\label{fig:quarter}
\end{figure*}

\paragraph*{Macrostates.} How can we picture the state of an oscillator network at a glance? Draw each
oscillator as a point on the unit circle at its phase, and join two oscillators by a chord whenever they
share an edge in the network. In this picture what stands out is not the individual oscillators but the
piles of oscillators that sit at the same phase, and the bundles of chords between piles. That suggests a
coarser description. Merge each pile into one dot, a \textit{cluster} $a$, at phase $\varphi_a$, with mass
$x_a$, the fraction of oscillators it contains, and replace each bundle of chords by one line, weighted by
the link density $P_{ab}=P_{ba}\in[0,1]$, the fraction of possible links that are present. What remains,
$T=(x,\varphi,P)$, is what we call the state's \textit{macrostate} (Fig.~\ref{fig:coarse}; a
\emph{template} in Ref.~\onlinecite{OKeeffeLB}). Every state has one, and it is far simpler than the network when oscillators pile up into a few clusters, as they do in the states that matter most here: the square states and their near relatives (Fig.~\ref{fig:quarter}).

The \textit{degree} of a macrostate,
\begin{equation*}
\mu(T)=\min_a\sum_bx_bP_{ab},
\end{equation*}
is the smallest cluster degree, measured as a fraction of all oscillators. For a macrostate formed from a
graph, each cluster's degree is the average degree of its oscillators divided by $n$, so $\mu(T)$ plays the role of the
connectivity. Counting each oscillator's harmless link to itself adds $1/n$ to every cluster degree, so minimum degree $\delta$ gives $\mu(T)\ge(\delta+1)/n$ (Appendix~\ref{app:setting}).

\paragraph*{Stability from the macrostate.} Macrostates have two benefits. The first is that the stability tests we need can be read off the macrostate, without the network: a stable state of the network gives a macrostate satisfying three necessary conditions on $(x,\varphi,P)$:
\begin{subequations}\label{eq:three}
\begin{gather}
\textstyle F_a=\sum_bx_bP_{ab}\sin(\varphi_a-\varphi_b)=0,\label{eq:Fa}\\
\textstyle c_a=\sum_bx_bP_{ab}\cos(\varphi_a-\varphi_b)\ge0,\label{eq:ca}\\
\textstyle Q(f)=\sum_{a,b}x_ax_bP_{ab}\cos(\varphi_a-\varphi_b)\,|f_a-f_b|^2\ge0.\label{eq:Qf}
\end{gather}
\end{subequations}
The first says that the pulls on each cluster cancel (equilibrium). The other two are the two halves of
stability, \textit{microscopic} and \textit{macroscopic}. The second concerns nudging a single oscillator away from its cluster: the net pull of its
neighbors, of strength $c_a$ on average, must point back toward the cluster's phase rather than away from
it, like a ball in a valley rather than on a hilltop. The third concerns moving whole clusters, that is, perturbing the macrostate itself: the energy
must not curve downward when each cluster $a$ is displaced by $f_a$. A complex $f$ describes two real displacement patterns at once, since
$Q(f)=Q(\mathrm{Re}\,f)+Q(\mathrm{Im}\,f)$, so to rule out a macrostate it is enough to exhibit one complex
vector $f$ with $Q(f)<0$. A macrostate that satisfies all three conditions is \textit{linearly stable}: it
is in equilibrium, and no small displacement lowers its energy to second order. Every stable state of a
network gives a linearly stable macrostate (Appendix~\ref{app:setting}).\footnote{Links inside a cluster drop out of the first and third
conditions, since $\sin0=0$ and $f_a-f_a=0$; they only add to the degree and to $c_a$, so completing them
keeps a macrostate linearly stable and can only raise its degree.}

\paragraph*{The size of the network drops out.} The second benefit of macrostates is that the number of oscillators
$n$ does not matter. Nothing in the three conditions, or in the degree, refers to $n$: only fractions of oscillators and
fractions of links appear. Graphs of every size, from a handful of nodes to millions, compress to the same
kind of object, so one argument about macrostates settles every graph at once. Despite the name, no large-$n$ limit is taken: the reduction is exact at any finite $n$. The macrostate retains the energy of rigid cluster motions exactly, while individual oscillators enter only through class averages. The price is that a
macrostate may have any number of clusters, with masses as small as we like, and the proof must work
uniformly over all of them.

\paragraph*{A key example of a macrostate: $C_4$.} $C_4$ is both the simplest example of a macrostate and
the one that causes all the trouble. It has masses $1/4$ at phases $0,\pi/2,\pi,3\pi/2$, with every link
present inside each cluster and between adjacent clusters, and none between opposite clusters
[Fig.~\ref{fig:quarter}(a)]. Each cluster is linked to three of the four, so $\mu(C_4)=3/4$. Checking the
conditions is easy: its two neighbors pull on it equally and oppositely, so $F_a=0$, and $c_a=1/4$. The
trouble is the third condition: $Q(f)=0$ for \emph{every} $f$, because links between adjacent clusters
carry the weight $\cos(\pi/2)=0$ and links inside a cluster have $f_a-f_a=0$. So $C_4$ is linearly stable
at degree $3/4$, yet its energy is completely flat to second order when its clusters are moved rigidly.
This is the razor's edge identified by Kassabov, Strogatz and Townsend~\cite{KST2021}, and the reason why $3/4$ was the old
barrier: an argument using only equilibrium and second-order conditions cannot exclude $C_4$ at degree $3/4$. We call it the \textit{blind
spot} of linear stability.

\paragraph*{The square states.} The blind spot is not special to $C_4$. Links between clusters a quarter
turn apart contribute $\cos(\pi/2)=0$ to $Q$: they add to the degree at no cost in stability. Links half a
turn apart contribute $\cos\pi=-1$ and destabilize, and links between clusters at nearly the same phase stabilize. So the dangerous
states are the square states of the Introduction (Fig.~\ref{fig:square}), whose phases all sit at the
corners of one square, $\theta_*+k\pi/2$ with $k=0,1,2,3$, with any masses at the corners.

The order parameters $\rho_k=\sum_ax_ae^{ik\varphi_a}$ detect them. Multiplying every phase by four sends
the four corners to a single point, so the square states are exactly the states with
\begin{equation*}
|\rho_4|=1:
\end{equation*}
their fourth harmonics are perfectly synchronized, just as $|\rho_1|=1$ means ordinary sync. The clusters
at or near each corner form the four \textit{groups}. To measure how far a macrostate is from the square
states we use
\begin{equation}
D_4=1-|\rho_4|^2\in[0,1],
\end{equation}
which vanishes exactly for them and equals $1$ when $\rho_4=0$, for example for five equally weighted, equally spaced phases. When every phase is within a small offset $u_a$ of a corner,
$D_4\approx16\,\mathrm{var}(u)$, where $\mathrm{var}(u)$ is the mass-weighted variance of the offsets. So
small $D_4$ means close to some square state, with arbitrary group masses; it does not mean close to
$C_4$.

\paragraph*{Two more important examples of macrostates.} Two more macrostates will accompany us through the
paper. The first [Fig.~\ref{fig:quarter}(b)] is a typical state near the square states: most of its mass
sits in main clusters, in thin groups at the corners, plus a few stray clusters elsewhere. Such states are
what Sec.~\ref{sec:shell} must rule out. The second is the family of split states $S_7(s)$ [Fig.~\ref{fig:quarter}(c)],
the best stable nonsynchronous states known~\cite{OKeeffeLB}: four corner clusters, three of them split into
two halves sitting a small angle $s$ on either side of their corner (the middle one a little wider). As the
splitting $s\to0$, their degree
approaches $\mu_0=0.6929\ldots$, which gives the lower bound $\mu_c\ge\mu_0$; at the same time they
approach the square states, with $D_4\to0$. So the states that come closest
to beating sync live near the square states. The member we plot, $S_7(0.002)$, has $\mu=0.6927$ and
$D_4=6.4\times10^{-5}$, still in the far region; the family enters the near region only for smaller $s$.

%==========================================================================================
\section{How the three regimes fit together}\label{sec:plan}

\begin{figure*}
\centering
\includegraphics[width=\textwidth]{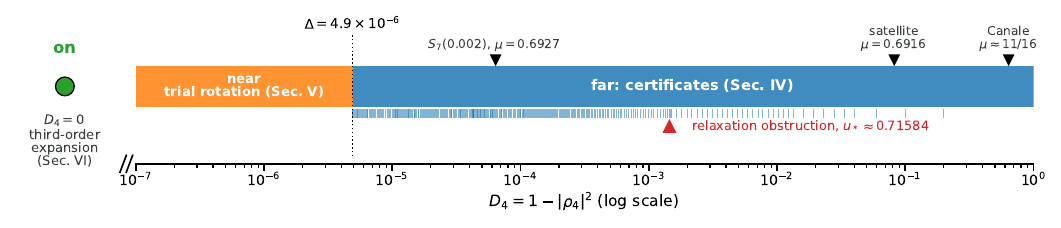}
\caption{The $D_4$ axis (log scale) and the three regimes, each with its own technique. Square states
($D_4=0$) sit at the left, past the axis break. The near region $0<D_4\le\Delta=4.9\times10^{-6}$ is
orange and continues below the plotted range to every positive $D_4$; the far region, covered by 310
certificate bands (ticks mark their left ends), is blue. Black triangles mark known examples, all with
degree below $0.717$: Canale's eight-cluster state, the previous record at $11/16$~\cite{Canale2022}
(linearly stable at that limiting degree, approached by stable networks), and two stable states from
Ref.~\onlinecite{OKeeffeLB}: the satellite state, in which two neighboring clusters of $C_4$ each shed a
small satellite cluster, and $S_7(0.002)$. The red triangle marks the feasible moment vector at
$D_4=0.00145$ that blocks the far relaxation up to threshold $0.71584\ldots$ (Sec.~\ref{sec:limits}).}
\label{fig:proof}
\end{figure*}

We can now make the plan of the Introduction precise. Theorem~\ref{thm:main} says that no
nonsynchronous stable state exists at high connectivity. Suppose one did. Its macrostate would be linearly
stable (Sec.~\ref{sec:setting}), with degree at least the target $\mus$, and it would have some distance
$D_4\in[0,1]$ from the square states. We rule out each possible value of $D_4$ in turn. The target degree
is set by the methods themselves: lowering it makes both the far and near arguments harder, and
$\mus=717/1000$ is about as low as we could push it while the two still meet. The cutoff between them is
$\Delta=49/10^7=4.9\times10^{-6}$, and we use it to split the unit interval:
\begin{equation}
[0,1]=[\Delta,1]\ \cup\ (0,\Delta]\ \cup\ \{0\}
\label{eq:partition}
\end{equation}
(Fig.~\ref{fig:proof}), and we work inward: Theorem~\ref{thm:far} in the \textit{far region}
(Sec.~\ref{sec:far}), Theorem~\ref{thm:shell} in the \textit{near region} (Sec.~\ref{sec:shell}), and
Lemma~\ref{lem:quarter} on the square states (Sec.~\ref{sec:square}). The two theorems overlap at
$\Delta$, so there is no gap, and the three results together prove Theorem~\ref{thm:main}
(Sec.~\ref{sec:proof}).

The two theorems reduce, by analytic arguments in the Appendixes, to finite lists of exact rational
checks: 310 certificates for the far region and 6486 boxes for the near region. The lemma is proved by
hand.

%==========================================================================================
\section{Far from the square states (large $D_4$): certificates}\label{sec:far}

We begin far from the square states, $D_4\ge\Delta$, with the most general tool. Here we must show that
no linearly stable macrostate has degree $\mus=0.717$ or more. The cluster phases $\varphi_a$ may be spread
anywhere around the circle, with no small parameter to expand in. So instead of hunting for a negative
eigenvalue state by state, we show that the conditions for linear stability and the degree bound cannot
all hold at once.

\begin{table}[t]
\caption{How each condition on a macrostate becomes a condition on its moments. Every row holds for every macrostate meeting the condition, whatever its number of clusters (Appendix~\ref{app:ledger}). PSD: positive semidefinite.}
\label{tab:moments}
\begin{ruledtabular}
\footnotesize
\begin{tabular}{@{}>{\raggedright\arraybackslash}p{0.27\columnwidth}>{\raggedright\arraybackslash}p{0.43\columnwidth}>{\raggedright\arraybackslash}p{0.24\columnwidth}@{}}
Condition & How it is used & Becomes\\
\hline
Equilibrium, Eq.~\eqref{eq:Fa} & weight by $x_ae^{ik\varphi_a}$, sum, multiply by $\rho_r\rho_s$ & linear equations (box)\\
Macroscopic stability, Eq.~\eqref{eq:Qf} & low-mode trial perturbations $f_a=e^{ik\varphi_a}\rho_r$ & PSD matrices (Hessian)\\
Microscopic stability, Eq.~\eqref{eq:ca} & $c_a$ as a nonnegative weight on each cluster & PSD matrices\\
Degree, $d_a\ge u$ & $d_a-u$ as a nonnegative weight on each cluster & PSD matrices\\
Links $0\le P_{ab}\le1$, masses $x_a\ge0$ & $\nu$, $\pi\otimes\pi-\nu$ and $\pi$ are nonnegative measures & PSD matrices (Gram)\\
Band, $D_4\in[\delta_-,\delta_+]$ & bounds on $|\rho_4|^2$, times $|\rho_j|^2$ & linear inequalities and PSD matrices\\
Order-parameter bounds & derived by hand from the macrostate conditions (Appendix~\ref{app:ledger}): $|\rho_1|\le1-u$, a quartic bound, a cap on $|\rho_2|$ & linear inequalities\\
\end{tabular}
\end{ruledtabular}
\end{table}

\begin{theorem}[Far region]\label{thm:far}
Every linearly stable macrostate with $D_4\ge\Delta$ has $\mu(T)<\mus$.
\end{theorem}

The argument has three steps. We move to Fourier space, where every condition becomes convex; we relax
the problem by treating the moments as unknowns; and we prove that the relaxed problem has no solution by
exhibiting a certificate. The first two steps are setup; the certificates are the heart.

\paragraph*{Fourier space.} A macrostate may have any number
of clusters, so there is no fixed list of unknowns to reason about. We therefore map each macrostate to its
Fourier moments, its averages, and translate every
condition we know into a condition on those moments: equilibrium, stability, the degree bound, the limits
on links, and bounds on the order parameters (Table~\ref{tab:moments}). Moment space has two virtues. Its
coordinates are the same however many clusters the macrostate has; there are infinitely many, but
keeping only the first few Fourier modes merely drops conditions, so every stable macrostate still
satisfies what is left. And the translated
conditions are convex: linear equations, linear inequalities, and positive semidefinite matrices. For
convex conditions, the statement ``no point satisfies them all'' can be proved by a certificate, which we
explain below.

Concretely, the phase measure $\pi=\sum_ax_a\delta_{\varphi_a}$ records where the clusters sit, and the
edge measure $\nu=\sum_{a,b}x_ax_bP_{ab}\delta_{(\varphi_a,\varphi_b)}$ records which clusters are
linked. Their Fourier moments, such as the order parameters $\rho_k=\int e^{ik\phi}\,\dd\pi$, are the
coordinates. The conditions are nonlinear in $(x,\varphi,P)$, but they become linear once we also keep
selected products of moments (for example, edge moments times order parameters) as variables of their own.
Equilibrium gives linear equations, and every inequality (stability, degree, the bounds on links and
masses) gives a matrix of moments that must be positive semidefinite, because a weighted average of
squares with nonnegative weights cannot be negative. The equilibrium box works out the first row of
Table~\ref{tab:moments}.

 \begin{figure}[t]
\begin{ideabox}[unbreakable,before upper={\leftskip=0pt\rightskip=0pt\parfillskip=0pt plus 1fil\parindent=0pt}]{A constraint, worked out: equilibrium}
Equilibrium, Eq.~\eqref{eq:Fa}, says $F_a=\sum_bP_{ab}x_b\sin(\varphi_a-\varphi_b)=0$ for every cluster
$a$. Weight it by $x_ae^{im\varphi_a}$ and sum over $a$. With $N_{ab}=x_ax_bP_{ab}$ and
$\sin\theta=(e^{i\theta}-e^{-i\theta})/2i$,
\begin{align*}
\sum_ax_ae^{im\varphi_a}F_a&=\frac1{2i}\sum_{a,b}N_{ab}\bigl(e^{i(m+1)\varphi_a-i\varphi_b}\\
&\qquad\qquad-e^{i(m-1)\varphi_a+i\varphi_b}\bigr)\\
&=\frac{1}{2i}\bigl(n_{m+1,-1}-n_{m-1,1}\bigr),
\end{align*}
where $n_{pq}=\sum_{a,b}N_{ab}e^{i(p\varphi_a+q\varphi_b)}=\int e^{i(p\phi+q\psi)}\dd\nu$ is a Fourier
moment of the edge measure. So $n_{m+1,-1}=n_{m-1,1}$ for every $m$. Multiply by order parameters
$\rho_r\rho_s$ with $r+s=-m$, so that the total frequency is zero and the product survives the average
over rotations. Writing $N(p,q;r,s)=\mathbb{E}[n_{pq}\rho_r\rho_s]$, with $\mathbb{E}$ the average over
rotations and a reflection,
\begin{equation*}
N(m+1,-1;r,s)-N(m-1,1;r,s)=0.
\end{equation*}
This is one linear equation in the moment variables, with integer coefficients, valid for every
macrostate in equilibrium, whatever the number of clusters.

\smallskip
The most important row of Table~\ref{tab:moments} comes from stability. Take any trial perturbation
built from the first few Fourier modes, $f_a=\sum_kv_k\,e^{ik\varphi_a}$. Then $Q(f)$ is a quadratic form
in the $v_k$ whose entries are edge moments, so $Q\ge0$ says that this matrix of moments is positive
semidefinite. Such a block tries every combination of its trial perturbations at once: the job the trial
rotation of Sec.~\ref{sec:shell} does by hand, one state at a time.
\end{ideabox}
\end{figure}

Two simplifications make the problem finite, and both are safe because they only weaken the conditions.
Rotating or reflecting a stable macrostate gives another one, so we may average its moments over all
rotations and a reflection. This kills every moment whose frequencies do not add to zero and makes the
rest real numbers in $[-1,1]$. And keeping only finitely many Fourier modes and products drops
conditions, so every stable macrostate still satisfies the rest. The cutoffs trade strength against
size, like the basis in a variational calculation; ours leave 238 variables (Appendix~\ref{app:far}).

\paragraph*{Relaxation.} We now have a finite list of convex conditions. Treat the moments as unknowns in their own right, subject only to these
conditions. The result is a \textit{relaxation}. Its feasible set contains the moments of every linearly
stable macrostate with degree at least $u$ and $D_4$ in a given range, together with impostors that come
from no macrostate. The logic runs in one direction, but it is the direction we need: if the relaxation
has no feasible point at $u=0.717$, no macrostate does, whatever its number of clusters.

\paragraph*{Certificates.}
This is the main step: we show that the relaxation has no solution. To prove that a list of conditions has no solution, combine them into a contradiction. If a
number $v$ had to satisfy $v-\tfrac12\ge0$ and $\tfrac14-v\ge0$, adding the two would give
$-\tfrac14\ge0$, which is impossible; the weights $(1,1)$ are the certificate. Finding such weights can be
hard, but once found, anyone can check them by adding. Ours combine several hundred conditions, some of them matrix inequalities, with
weights found by an optimization solver and then verified in exact rational arithmetic.
A \textit{dual certificate} is the same idea for the relaxation. In the toy the variable cancels
completely; in ours a little of each moment may be left over, and the bound $|v_k|\le1$ disposes of it.
The certificate consists of nonnegative weights $\lambda_j$ for the inequalities $g_j(v)\ge0$, weights
$\kappa_i$ of either sign for the equations $h_i(v)=0$, and positive semidefinite matrix weights $Y_l$ for
the matrix conditions $H_l(v)\succeq0$, such that
\begin{multline}
\sum_j\lambda_j\,g_j(v)+\sum_i\kappa_i\,h_i(v)+\sum_l\operatorname{tr}\bigl(Y_lH_l(v)\bigr)\\
=b_0+\sum_kr_kv_k,\qquad b_0+\sum_k|r_k|<0 .
\end{multline}
At a feasible point the left side is nonnegative, but the right side is negative, since every moment lies
in $[-1,1]$. So there is no feasible point (Appendix~\ref{app:far}).

\paragraph*{Bands.}
One refinement makes this work in practice. A single relaxation for the whole far region would be too weak, because it would have to accommodate states
close to the blind spot. So we cut $[\Delta,1]$ into bands and give each band its own relaxation, with
the band's limits on $D_4$ added as constraints, and its own certificate. Next to the near region the relaxation is weakest, so the bands there are thinnest: 135 of the 310 bands
lie below $D_4=5.1\times10^{-5}$, while a single band covers $[0.2,1]$ (Table~\ref{tab:cover}). Each
band's certificate excludes every degree $u$ or more, for some $u\le0.717$ (306 bands use $u=0.717$
itself), so together they exclude every linearly stable macrostate in the far region with degree at least $0.717$.

\paragraph*{Is the relaxation valid?}\label{sec:ledger} Finally, what we regard as the weakest link. A certificate refutes the relaxation, not the
macrostates directly. So everything rests on one claim: \emph{every} linearly stable macrostate with
$\mu(T)\ge u$ and $D_4$ in the band satisfies every constraint. If one constraint did not follow from
linear stability, every certificate could be correct as arithmetic and still prove nothing.
Appendix~\ref{app:ledger} derives each constraint family, and the box at the end of
Sec.~\ref{sec:proof} describes how we tested them on genuine macrostates.

\paragraph*{How low can certificates go?}\label{sec:limits} Not much lower: an obstruction shows that
$0.717$ is close to the end of this method. At $D_4=29/20000=0.00145$ (the red triangle in
Fig.~\ref{fig:proof}) there is an explicit rational point that satisfies every constraint of the far
relaxation, exactly, at every threshold $1/2<u\le22906937/32000000=0.71584178125$ and on every band that
contains $0.00145$ (Proposition~\ref{prop:appb-obstruction}). So no certificate exists there, and no
subdivision in $D_4$ can remove the point. This is a limit of the relaxation, not a counterexample: the
moments need not come from any macrostate, and $0.71584\ldots$ is still far from $\mu_0=0.6929$. Extra
moment constraints, conditioning on the size of $\rho_2$, or an analytic estimate reaching
$D_4=0.00145$ could bypass it.

\paragraph*{Why the certificates stop at $\Delta$.} Certificates cannot reach the square states. At
$D_4=0$ there is nothing to contradict: $C_4$ has degree $3/4>\mus$ and is linearly stable, so its
moments satisfy every constraint of the relaxation except the band's limits on $D_4$. As a band
approaches $D_4=0$, the contradiction therefore becomes vanishingly weak, which is why the bands there are
so thin. This forces us to stop at $\Delta$ and switch to a new method that uses the geometry of the
corners: the trial rotation.

%==========================================================================================
\section{Near the square states (small $D_4$): the trial rotation}\label{sec:shell}

Next comes the near region, $0<D_4\le\Delta$, the heart of this paper. States in this region look like
slightly blurred square states: tight groups of clusters near the corners, plus a few stragglers farther away
[Fig.~\ref{fig:quarter}(b)]. Our claim is that once the degree reaches $\mus$, every one of them is
unstable, with a negative eigenvalue, so none of them can rival sync as a stable state. Certificates
cannot show this, as we saw: these states are too close to $C_4$. But blurring lifts the zero eigenvalues of the square states, so the sign is decided at second order, and the question is whether a lifted
eigenvalue goes negative. So we return to individual states and use the variational principle: one explicit rule builds, for each of these states, a perturbation $f$ that lowers its energy, $Q(f)<0$.

\begin{theorem}[Near region]\label{thm:shell}
There is no linearly stable macrostate with $\mu(T)\ge\mus$ and $0<D_4\le\Delta$. Indeed, for any
macrostate with these properties we construct a complex vector $f$ with
\begin{equation}
Q(f)\le-10^{-10}D_4<0,
\label{eq:shellQ}
\end{equation}
contradicting linear stability.
\end{theorem}

\paragraph*{What makes the near region hard.} The negative eigenvalue we are hunting is tiny. Just off the
square states the negative curvature we can establish is only of order $D_4$, which can be arbitrarily
small; that is why Eq.~\eqref{eq:shellQ} is linear in $D_4$. The trial rotation must find this sliver of negative curvature uniformly,
for macrostates with any number of clusters and arbitrarily small masses.

\paragraph*{The strategy} uses stability against itself. We assume a state in the near region is linearly
stable, use simple perturbations to learn its shape (Steps~1--2), and then build a trial rotation that this
shape cannot survive (Steps~3--4). The plan is that of Lepsveridze and Zhang (Ref.~\onlinecite{LZ2026},
Secs.~4--5): rotate the main clusters together, and give each stray cluster its own perturbation. What is
new is that every step is explicit. That takes a quadratic correction to the rotation (Step~3) and, for
the stray clusters, a finite list of computer checks that together cover every possible stray cluster
(Step~4). The checks use exact arithmetic with fractions, never rounded decimals, so each one that passes
is a proof. The complete derivation, with all
constants (Table~\ref{tab:constants}), is in Appendix~\ref{app:shell}.

\paragraph*{Step 1: main and stray clusters.}
Rotate so that $\rho_4=r\ge0$; since $D_4=1-r^2$ is tiny, $r\ge r_0=999/1000$ throughout the near region.
Write $\varphi_a=j(a)\pi/2+u_a$ with $|u_a|\le\pi/4$: cluster $a$ belongs to group $j(a)$ and sits at
offset $u_a$ from its corner. Then the exact identity
\begin{equation}
\sum_ax_a\sin^2(2u_a)=\frac{1-r}{2}=\frac{D_4}{2(1+r)}\le S_*\equiv\frac{\Delta}{2(1+r_0)}
\label{eq:S1}
\end{equation}
can be read as a budget: each cluster pays
$\sin^2(2u_a)$ per unit mass, and the whole macrostate can afford only $S_*$, of order $\Delta$. Fix a cutoff angle $h=483/5000\approx0.1$ rad. The \emph{main} clusters of group $j$, $\B_j=\{a:j(a)=j,\ |u_a|\le h\}$, sit within $h$ of their corner; the rest, the set $\E$, are \emph{stray}. Write $x(S)=\sum_{a\in S}x_a$ for the mass of a set $S$ of clusters, and $X_j=x(\B_j)$ for the main mass of group $j$. A stray cluster pays at
least $\sin^2(2h)$ per unit mass, so the stray mass $b=x(\E)$ is small. A main cluster pays about $4u_a^2$,
so the second moment $\sum_\B xu^2$ of the main clusters is small. Both bounds are of order $\Delta$
(Table~\ref{tab:constants}). Stray clusters are the nuisance: their total mass is tiny, but it can be as large as a constant times $D_4$, so it can compete with the curvature we are trying to establish.

\paragraph*{Step 2: all four groups are occupied, nearly equally.}
Think of the four main group masses $X_j$ (the mass near each corner) as a signal on the four corners. Up to small corrections, its
discrete Fourier transform is the set of order parameters $\rho_0,\dots,\rho_3$. Stability keeps $|\rho_1|$
and $|\rho_2|$ small (Lemma~\ref{lem:order}), so no group can be empty, and near-sync, two-group and
three-group states are excluded at once. Two more facts sharpen this. The stability test that moves one group as a block, combined with torque balance, shows that opposite groups are almost unlinked, and equilibrium makes the links a
group sends to its two neighbors almost balance. Together these pin every $X_j$ near $1/4$ (between about
$0.22$ and $0.28$), and give
\begin{equation}
X_j+2X_i\ge\mus-\varepsilon,\qquad \varepsilon\approx1.8\times10^{-4},
\label{eq:S2}
\end{equation}
for every pair of adjacent groups $j,i$ (Appendix~\ref{app:shell}). For $C_4$ the left side is $3/4$. So the
blurred square state is in fact close to $C_4$: all four corners are filled, nearly equally, plus a little
stray mass.

\begin{figure}[t]
\centering
\includegraphics[width=\columnwidth]{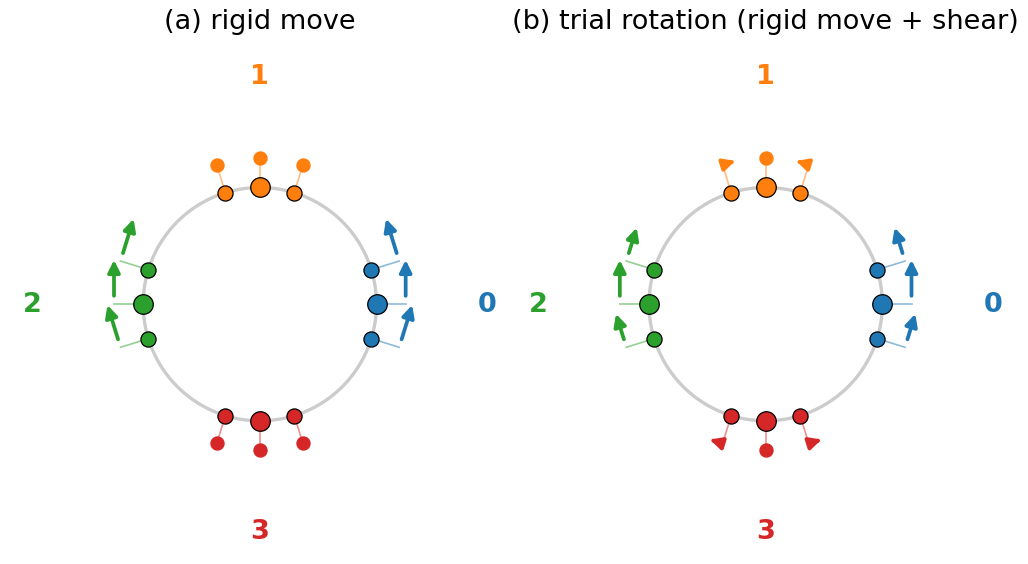}
\caption{The trial rotation on a blurred square state (offsets exaggerated; arrows show phase
displacements). (a) Rigid move: groups 0 and 2 counter-rotate, flat at $C_4$. (b) Adding the shear:
clusters of groups 1 and 3 move in proportion to their offsets, with slopes $-k$ and $+k$. Shown is the
real part of $f$ in Eq.~\eqref{eq:trial}.}
\label{fig:trial}
\end{figure}

\needspace{5\baselineskip}
\paragraph*{Step 3: the trial rotation.}
Start from $C_4$, where every rigid cluster displacement is flat to second order (Sec.~\ref{sec:setting}), and pick a useful one: rotate
group $0$ forward and group $2$ backward by the same small angle, and leave groups $1$ and $3$ alone [Fig.~\ref{fig:trial}(a)]. The
move keeps each group rigid, so the stabilizing links inside groups are not stretched. It also pulls
opposite groups apart as hard as it can, so any destabilizing links between them count in full. Off the
square states, though, the move is not reliably downhill. Each group now has a small spread of offsets
$t_a=u_a-\bar u$, measured from the mean offset $\bar u$ of the main clusters, and a link between a
cluster $a$ in group $j$ and a cluster $b$ in group $j+1$ enters $Q$ with the weight $\sin(t_a-t_b)$,
which is small and can have either sign.

The remedy is a shear that reads those signs [Fig.~\ref{fig:trial}(b)]. While groups $0$ and $2$ counter-rotate, groups $1$ and $3$
are sheared in opposite directions: each of their clusters moves by an amount proportional to its offset
$t_a$, with slope $-k$ in group $1$ and $+k$ in group $3$ ($k=1.377$, Table~\ref{tab:constants}). Across every adjacent link with positive weight
the shear reduces the relative displacement, and across every link with negative weight it increases it.
Its leading contribution to $Q$, proportional to $-k(t_a-t_b)\sin(t_a-t_b)$, is therefore nonpositive on every adjacent link, and strictly negative when the offsets differ. The same move with the roles of the two pairs exchanged is used as well. In complex notation both
moves fit into one vector, whose real and imaginary parts they are:
\begin{equation}
f_a=i^{j(a)}F(t_a),\qquad F(t)=1+\ell t^2+ikt.
\label{eq:trial}
\end{equation}
The small term $\ell t^2$, with $\ell=-1-k^2/2$, slightly shortens the rigid part. This is the
\emph{quadratic correction}. The rigid part leaves behind a term of either sign on every adjacent link, and
$\ell$ is tuned so that the equilibrium equations cancel that term up to small remainders.

What is left is a contest between links. Adjacent links that are present make $Q$ more negative, through
the shear. Links inside a group make it more positive, because the shear moves their two ends by different
amounts. The negative side wins if few adjacent links are missing, and the degree bound guarantees this: a
cluster of degree at least $\mus$ cannot miss many of its adjacent links. Making this quantitative,
uniformly over all macrostates, is the technical core of Appendix~\ref{app:shell}, which counts missing
links group by group with weights tuned to the group masses. The outcome is that the main clusters
contribute to $Q(f)$ at most a fixed negative multiple of their spread $\sigma^2=\sum_\B x_at_a^2$
(denoted $V$ in Appendix~\ref{app:shell}), plus a boundary term from their links to stray clusters.

\paragraph*{Step 4: stray clusters pay their own way.}
Each stray cluster gets its own unit complex number in $f$, chosen from its phase and from how strongly it
is linked to the four main groups. To make its contribution to $Q(f)$ computable, idealize: place the main
clusters at their corners, and adjust the stray cluster's links slightly so that its torque balances
exactly. The idealized contribution is then bounded in terms of two numbers: the cluster's angle from the nearest corner and the size of the imbalance of its links to opposite groups; the remaining link masses are bounded within each box. We cover that two-parameter rectangle by $6486$
boxes. On each box an exact rational check shows either that no stray cluster fits there or that the
idealized contribution is negative throughout. Undoing the idealization costs a little, and links between
stray clusters can add positive curvature. Summed over all stray clusters, these costs are absorbed by the
negative terms, with no case analysis on how many stray clusters there are or where they sit. Adding everything gives
$Q(f)\le-10^{-10}(16\sigma^2+4b)$, where $b$ is the stray mass. Since $D_4\le16\sigma^2+4b$, this is
Eq.~\eqref{eq:shellQ} (Appendix~\ref{app:absorb}).

The guarantee of Eq.~\eqref{eq:shellQ} is proportional to $D_4$, so it fades as a state approaches the
square states and vanishes on them. There the second-order test is truly blind, and we need third order
(Sec.~\ref{sec:square}).

%==========================================================================================
\section{On the square states ($D_4=0$)}\label{sec:square}

Finally, the square states themselves. No second-order argument, certificates or trial rotations alike, can show that one with links between groups is unstable, since $C_4$ passes
every second-order test at degree $3/4$. So we go to third order. A nonnegative Hessian is not enough here, since it permits the cubic descent below; this is the one place where the proof uses a genuine local minimum rather than linear stability. The result, due to
Lepsveridze and Zhang, is that on a dense graph the only stable square state is sync.

\begin{lemma}[Links in square-state minima~\cite{LZ2026}]\label{lem:quarter}
Let $\theta$ be a local minimum of the energy $V$ of Eq.~\eqref{eq:potential} whose phases lie at the corners of one square, $\theta_*+k\pi/2$ with
$k=0,1,2,3$. Then every
edge of $G$ joins two vertices with the same phase.
\end{lemma}
In words: a stable square state has no links between different groups. A graph with connectivity
$\mu>1/2$ is connected, since any two oscillators are linked or share a neighbor, so all oscillators then sit in one
group, which is sync.

Their proof~\cite{LZ2026} is short (Appendix~\ref{app:setting}). Number the groups
$0,1,2,3$ around the circle. A stable square state has no links between opposite
groups, since such links carry the weight $\cos\pi=-1$ and would destabilize it. And equilibrium forces the
same number of links, $m$, between each pair of adjacent groups, as in $C_4$. Now rotate group $0$ by a
small angle $t$ and group $1$ by $-t$. Each link between groups $0$ and $1$ changes its energy by
$-\sin2t$, and each link from these two groups to their other neighbors changes it by $+\sin t$. The total
is
\begin{equation*}
\delta V=m\,(2\sin t-\sin2t)\approx mt^3.
\end{equation*}
The terms of order $t$ and $t^2$ cancel, as they must at the blind spot, but the cubic term has a sign: for
small $t<0$ the energy drops. So if $m>0$, $C_4$ and every such square equilibrium, whatever its group masses, is a degenerate saddle, not a minimum. If $m=0$, every edge joins equal phases, and the degree hypothesis forces sync.

%==========================================================================================
\section{Proof of the main result}\label{sec:proof}

Suppose a graph with $(\delta+1)/n\ge\mus$ had a nonsynchronous stable state. Its
macrostate, with self-links added, is linearly stable with degree at least $\mus$ (Sec.~\ref{sec:setting}).
If $D_4\ge\Delta$, Theorem~\ref{thm:far} says its degree is below $\mus$; if $0<D_4\le\Delta$,
Theorem~\ref{thm:shell} says it is not linearly stable. So $D_4=0$: the oscillators sit at the corners of
one square, and Lemma~\ref{lem:quarter} says every link joins equal phases. A graph this dense is
connected (any two oscillators are linked or share a neighbor), so the state is synchronized, a
contradiction.~$\square$ (Details, including $n\le2$, are in Appendix~\ref{app:setting}.)

\begin{ideabox}[before upper={\leftskip=0pt\rightskip=0pt\parfillskip=0pt plus 1fil\parindent=0pt}]{How sure are we?}\label{sec:margins}
The proof leans on computation in two places: the 310 certificates of the far region and the 6486 boxes
of the near region. No part of the proof has been formalized in a proof assistant.

\textbf{Is the arithmetic right?} Every check is exact rational arithmetic in short standard-library
Python programs (solvers only found certificates), available from the author and independently repeated (Methods).

\textbf{Do the checks prove the theorems?} Only through the analytic arguments that link them to
macrostates: Appendix~\ref{app:shell} for the near region, and the validity of the relaxation
(Appendix~\ref{app:ledger}) for the far region. We also tested the latter: independently rebuilt constraints matched the originals, and every
constraint held on 369 validated stable macrostates covering every band (Methods).

\textbf{How much room is there?} Very little: every near-region check passes with an exact positive
margin, but some margins are tiny (relative slack $1.2\times10^{-7}$ in the box checks,
Table~\ref{tab:margins}).
\end{ideabox}

%==========================================================================================
\section{Discussion}\label{sec:discussion}

\paragraph*{Where is $\mu_c$?} We now know $0.6929077\ldots\le\mu_c\le0.717$. The best stable states we
know hug the square states and approach the lower end~\cite{OKeeffeLB}; the upper bound is limited by the
method.

\begin{figure}[t]
\begin{conjbox}[before upper={\leftskip=0pt\rightskip=0pt\parfillskip=0pt plus 1fil\parindent=0pt}]
The critical connectivity is
\begin{equation}
\mu_c=\mu_0=\frac{4+2q_*}{7+q_*}=0.6929077\ldots,
\label{eq:mu0}
\end{equation}
where $q_*\approx0.6506$ is the unique root in $(0,1)$ of
\begin{multline}
9q^8+72q^7+288q^6+448q^5+470q^4\\
+88q^3-208q^2-128q-15=0.
\label{eq:qstar}
\end{multline}
\end{conjbox}
\end{figure}

\paragraph*{Other methods.} The natural alternative is the landscape of the Burer--Monteiro
factorization: for a connected graph whose Laplacian eigenvalues satisfy $\lambda_n/\lambda_2<2$, every
second-order critical point of $V$ is synchronous~\cite{RakotoEndor2024,McRae2025}. Combined with
Fiedler's bound $\lambda_2\ge2\delta-n+2$ and $\lambda_n\le n$, this recovers the threshold $3/4$. But networks
built on $C_4$ approach that degree with $\lambda_n/\lambda_2=2$ exactly, so this criterion cannot give a
minimum-degree bound below $3/4$. Going lower takes information about the states themselves.

\paragraph*{Near the square states.} In work in progress, we rescale the offsets of a near-square macrostate by their overall size $s$ and let
$s\to0$; the leading-order problem is a finite-dimensional semidefinite problem. The evidence suggests that
its value is exactly $\mu_0$ and that the error at small positive $s$ can be controlled uniformly over
macrostates. If both hold, $\mu_0$ is sharp near the square states: for each $\epsilon>0$, every nonsynchronous
stable state with connectivity at least $\mu_0+\epsilon$ has $D_4\ge d_\epsilon>0$, while $S_7(s)$
approaches $\mu_0$. None of this is used in the
present paper.

\paragraph*{A conjecture.} Every nonsynchronous stable state we have found has degree below $\mu_0$, the
degrees of $S_7(s)$ approach it, and the evidence above points to $\mu_0$ near the square states. So we
conjecture that $\mu_0$ is the exact threshold (box; $q_*$ is the limiting density of the thinned links
in $S_7(s)$~\cite{OKeeffeLB}). A proof would also need the far region, where the present relaxation stalls at $0.7158$ (Sec.~\ref{sec:limits}).

\paragraph*{Outlook.} Two questions remain. First, dynamics:
does almost every initial condition synchronize above $\mu_c$? Degenerate saddles such as $C_4$ could
in principle attract a set of positive measure (as $\dot x=x^2$ does), so this needs their local
dynamics; here $3/4$ is still the best bound.
Second, does the same strategy work for other networks with such blind spots?

%==========================================================================================
\section*{Methods}\label{sec:ai}
\subsection*{Use of AI tools}

AI agents (Claude Opus~5.5 and Fable~5.1, Anthropic; GPT-6.1-sol, GPT-6.2-sol and GPT-6-astra, OpenAI;
Grok~4.7, xAI), guided by the author in all of these tasks, did much of this work: they developed the quantitative near-region estimates, wrote the near-region checker and the far-region certificate generators and checkers, found the far-region certificates, and drafted and revised the manuscript. Every
bound reported here was audited adversarially by at least two different model families, and every
certificate is checked in exact rational arithmetic by the verification programs (see Independent checks).
The author chose the problem, set the standard of evidence, decided which results to trust, and
edited the text, and takes full responsibility for the content.

\subsection*{Independent checks}

The near-region inequalities and box cover were checked by the original checker, by a second checker written separately,
and by two independent rechecks that recompute every constant and every box.
The far cover was checked by the original verifier, and every one of the 310 bands was re-summed
from its dual data by two independent programs. The audits looked for
errors in the arguments as well as in the code.

The far relaxation itself was tested as follows. An AI agent (GPT-6.1-sol, via
Codex) wrote the ledger of derivations behind Appendix~\ref{app:ledger} and built the constraints a second
time from its formulas, without importing the original code. On all 310 band specifications and ten
small variants the two builders agree coefficient by coefficient: 4{,}552{,}730 expressions, with zero
mismatches. A second model (Grok~4.7) audited the ledger, and its own builder also matches exactly. Every constraint was then evaluated, at 100 decimal digits with tolerance $10^{-65}$, on the moments of
369 independently validated linearly stable macrostates, each with its own degree as the threshold and in a
band containing its $D_4$. The controls include at least one in each of the 310 released bands, as well as
examples in additional positive bands below $\Delta$. They comprise 363 exact rational macrostates; four
macrostates validated by interval equilibrium and reduced-Hessian checks (two $S_7$ members and the
satellite from Ref.~\onlinecite{OKeeffeLB}, and REC4, an earlier near-square record state from our
searches with $\mu=0.6922549\ldots$); and two whose linear stability holds by exact algebra (Canale's
state and one state with $D_4=1$). They span $D_4$ from $1.6\times10^{-23}$ to $1$. Because most rational controls are two-group states near the corners,
20 further exact macrostates with other supports (split states with 12 and 16 clusters, tiny extra masses,
clusters moved off their group by up to $0.1$ rad) were added. All passed.

\section*{Data availability}
The verification code and certificates are available from the author on request. They comprise the frozen tree \texttt{release\_0717/} used for Theorem~\ref{thm:main}:
a README, \texttt{run\_checks.sh}, \texttt{SHA256SUMS}, the near-region data \texttt{shell\_717.json}
with its checker \texttt{check\_final.py}, the 310-leaf far cover under \texttt{far/certificates\_v4/}
with the verifier \texttt{far/verify4.py}, its helper modules and the constraint builder
\texttt{far/fourier2.py}, and the exact feasible point of Sec.~\ref{sec:limits}
(\texttt{primal\_peak.json}, with its recorded checks in \texttt{EXACT\_PEAK.json}). The script
\texttt{run\_checks.sh} reruns the near-region and far-cover checks and verifies the file hashes. Run it
from an extracted copy of \texttt{release\_0717/}, as described in Appendix~\ref{app:repro}; to run the
near-region checker alone, pass \texttt{shell\_717.json} to it explicitly. Next to the frozen tree, in the same package, \texttt{extras/appb\_checks.py} replays the exact feasible point of Sec.~\ref{sec:limits} and the supplementary identities (run \texttt{python3 -B appb\_checks.py} from \texttt{extras/}; it requires SymPy). The audit logs and the search code are also available on request.

\section*{Author Declarations}
\subsection*{Conflict of Interest}
The author has no conflicts to disclose.

\appendix

\section{Macrostates and the main proof}\label{app:setting}

At a local minimum the right-hand side of Eq.~\eqref{eq:kuramoto} vanishes and the Hessian
$H\succeq0$ has off-diagonal entries
$H_{jk}=-A_{jk}\cos(\theta_j-\theta_k)$ and zero row sums.
Only Lemma~\ref{lem:quarter} uses the local minimum beyond these conditions.

\subsection{Macrostates and aggregation}

\begin{definition}
A \emph{macrostate} $T=(x,\varphi,P)$ has finitely many clusters $a$, with masses $x_a>0$ summing to one,
phases $\varphi_a$, and symmetric densities $P_{ab}=P_{ba}\in[0,1]$, the diagonal included.
\end{definition}
The number of clusters is unbounded, and there is no positive lower bound on individual masses. Write $N_{ab}=x_ax_bP_{ab}$ and
\begin{align*}
 d_a&=\sum_bx_bP_{ab},& \mu(T)&=\min_a d_a,\\
 F_a&=\sum_bx_bP_{ab}\sin(\varphi_a-\varphi_b),\\
 c_a&=\sum_bx_bP_{ab}\cos(\varphi_a-\varphi_b).
\end{align*}
The complex quadratic form, summed over ordered pairs, is
\begin{equation}
 Q(f)=\sum_{a,b}N_{ab}\cos(\varphi_a-\varphi_b)|f_a-f_b|^2,
 \qquad f\in\mathbb C^{\text{clusters}}.
 \label{eq:Q}
\end{equation}
It equals $\mathcal L(\operatorname{Re}f)+\mathcal L(\operatorname{Im}f)$ for the corresponding real Laplacian form $\mathcal L$.
A macrostate is \emph{linearly stable} when $F_a=0$, $c_a\ge0$ for every $a$, and $Q\ge0$.

\begin{lemma}[Aggregation~\cite{OKeeffeLB}]\label{lem:agg}
Let $\theta$ be a fixed point of $G$ with $H\succeq0$. Group equal phases into classes $C_a$, and set
$x_a=|C_a|/n$, $\varphi_a$ the common phase, and $P_{ab}$ the number of ordered adjacent pairs between
$C_a$ and $C_b$ divided by $|C_a||C_b|$. This is a linearly stable macrostate with $\mu(T)\ge\delta/n$ and
$P_{aa}\le1-1/|C_a|$.
\end{lemma}

\begin{proof}
The class averages of graph degree, equilibrium sum and Hessian diagonal, divided by $n$, give $d_a,F_a,c_a$. Hence $F_a=0$ because $\theta$ is a fixed point, and $c_a\ge0$ because the diagonal of $H\succeq0$ is nonnegative.
On class-constant vectors $Q$ is the graph Hessian form on the real and imaginary parts scaled by the positive factor $2/n^2$, using ordered pairs, so $Q\ge0$ follows from $H\succeq0$. The absence of self-loops gives the diagonal density bound.
\end{proof}

Adding $1/|C_a|$ to each $P_{aa}$ keeps it at most one and leaves $F_a,Q$ unchanged since $\sin0=0$ and $f_a-f_a=0$. This adds $x_a/|C_a|=1/n$ to $c_a$ and $d_a$. For $n>1$, the resulting linearly stable $T'$ satisfies
\begin{equation}
 \mu(T')\ge(\delta+1)/n\ge\delta/(n-1).
 \label{eq:selfloop}
\end{equation}
The difference on the right is $(n-1-\delta)/(n(n-1))\ge0$.

\subsection{Square states and order bounds}

For $\rho_k=\sum_ax_ae^{ik\varphi_a}$, set $D_4=1-|\rho_4|^2\in[0,1]$.
Equality $D_4=0$ means all unit numbers $e^{4i\varphi_a}$ coincide, so the phases lie at the corners of one square,
$\theta_*+k\pi/2$ with $k=0,1,2,3$. These \emph{square} macrostates include sync, two- and three-group states, and $C_4$.
Small $D_4$ measures concentration near the corners of such a square and allows arbitrary group masses.

\begin{lemma}\label{lem:order}
Let $T$ be linearly stable with $\mu(T)\ge u>1/2$ and $D_4>0$. Put $a_0=1-u$, $\eta=3/4-u$,
$p=|\rho_1|^2$ and $y=|\rho_2|$. Then $p\le a_0^2$ and
\begin{equation}
2p+A_0y^2\le B_0,
\label{eq:order}
\end{equation}
where $A_0=\tfrac54-6\eta-12\eta^2$ and $B_0=8\eta+16\eta^2$.
\end{lemma}

\begin{proof}
In the fractional version of the argument of Ref.~\onlinecite{KST2021}, equilibrium gives
\[
 \rho_1-c_ae^{i\varphi_a}=\sum_bx_b(1-P_{ab})e^{i\varphi_b},
 \qquad |\rho_1-c_ae^{i\varphi_a}|\le a_0.
\]
If $|\rho_1|>a_0$, rotate to $\rho_1>0$. Then $c_a\cos\varphi_a>0$, so $c_a\ge0$ puts every phase in an open semicircle.
Every component of the positive-density graph has mass at least $u>1/2$, so there is only one component and the graph is connected.
At a maximal phase every sine in $F_a$ is nonnegative and strictly positive on edges to lower phases.
Thus the maximal set has no outside edge, forcing sync and contradicting $D_4>0$.

Testing $Q$ on $\cos\varphi,\sin\varphi$ gives $\sum_{a,b}N_{ab}(t_{ab}-t_{ab}^2)\ge0$ with $t_{ab}=\cos(\varphi_a-\varphi_b)$.
As derived in Appendix~\ref{app:ledger}, equilibrium, two weighted Cauchy--Schwarz bounds and $t-t^2\ge-2|t|$ on $[-1,1]$ give
\[
 0<1-2p+y^2\le\sqrt{16a_0^2-8p(1-y)}.
\]
Squaring gives Eq.~\eqref{eq:O2}. Using $8y\le2+8y^2$ bounds its left side below by
$1+2p+(2-12p)y^2+4p^2+y^4$.
Now use $p\le a_0^2$, $A_0=2-12a_0^2$ and $B_0=16a_0^2-1$.
\end{proof}

\subsection{Exact square states and the main proof}

\begin{proof}[Proof of Lemma~\ref{lem:quarter}]
For $B_j=\{v:\theta_v=\theta_*+j\pi/2\}$, writing $e(B_i,B_j)$ for the number of edges joining two groups, the Hessian test on $1_{B_j}$ gives $-e(B_j,B_{j+2})\ge0$.
Opposite cuts are therefore empty. Equilibrium equates the counts of neighbors a quarter turn ahead and behind at each vertex. Summing on each group makes all four adjacent edge counts equal to $m$.
Rotating $B_0$ by $t$ and $B_1$ by $-t$ leaves same-group energies and the $B_2$--$B_3$ cut unchanged.
The other adjacent cuts start at energy $1$ per edge, with changes $-\sin2t$ on $B_0$--$B_1$ and $\sin t$ on each of $B_0$--$B_3$ and $B_1$--$B_2$.
The total change $m(2\sin t-\sin2t)=2m\sin t(1-\cos t)$ is negative for small $t<0$ unless $m=0$.
\end{proof}

\begin{proof}[Proof of Theorem~\ref{thm:main}]
For $n=1$ the claim is immediate. Aggregate a local minimum and add the self-loops to obtain a linearly stable $T'$ with the same $D_4$ and $\mu(T')\ge(\delta+1)/n\ge\mus$.
Theorems~\ref{thm:far} and~\ref{thm:shell} exclude $D_4\ge\Delta$ and $0<D_4\le\Delta$, respectively, with overlap at $\Delta$.
At $D_4=0$, Lemma~\ref{lem:quarter} makes phases equal along every edge.
For $n\ge3$, $\delta\ge0.717n-1\ge(n-1)/2$, so nonadjacent vertices have a common neighbor by counting their neighbors among the other $n-2$ vertices.
For $n=2$, $\delta\ge1$ gives $G=K_2$. Connectivity therefore forces sync.
\end{proof}

\section{The far relaxation and its certificates}\label{app:far}

\subsection{A Fourier moment relaxation}

Let $T$ be linearly stable with $\mu(T)\ge u$ and $D_4$ in a closed band $[a,b]\subset(0,1]$. Write $\pi=\sum_ax_a\delta_{\varphi_a}$ and $\nu=\sum_{a,b}N_{ab}\delta_{(\varphi_a,\varphi_b)}$, so $\nu$ is symmetric and $0\le\nu\le\pi\otimes\pi$. Let $\rho_k=\int e^{ik\phi}\,\dd\pi$, the finite sum $\sum_ax_ae^{ik\varphi_a}$ of Appendix~\ref{app:setting}, and $n_{ab}=\int e^{i(a\phi+b\psi)}\,\dd\nu$. Let $\mathbb{E}$ be the average over a uniform rotation and an independent reflection of probability $1/2$. The variables
\begin{align}
X(k_1,\dots,k_t)&=\mathbb{E}\Bigl[\prod_i\rho_{k_i}\Bigr],\nonumber\\
N(a,b;k_1,\dots)&=\mathbb{E}\Bigl[n_{ab}\prod_i\rho_{k_i}\Bigr],
\end{align}
are joint moments of one macrostate, not products of averaged moments. Rotation deletes every monomial whose frequencies do not sum to zero, and reflection makes the survivors real. Each lies in $[-1,1]$: $\pi$ has mass one and $\nu$ has mass at most one.

Appendix~\ref{app:ledger} derives the families emitted by \texttt{far/fourier2.py}. They are affine in these moments, with rational coefficients. At the release cutoffs $(R,H,D,A,L)=(4,5,4,7,8)$, with symmetric edge blocks, conditioning modes $-R,\ldots,R$, and unconditional input $\rho_2$ band $[0,1]$, the builder returns $238$ variables and $40$ equalities. A $1\times1$ block is stored as a scalar row. When the cap of Appendix~\ref{app:ledger} is strictly below $1$, there are $513$ inequalities and $87$ PSD blocks if the upper endpoint is below $1$, and $507$ inequalities and $86$ PSD blocks if it equals $1$. The difference is five omitted scalar rows and one omitted localizer for the upper bound on $D_4$ (named ``lower exclusion'' in the builder, since it is a lower bound on $z=1-D_4$); the band product then duplicates the surviving scalar row at $j=4$ and is stored once. Of the $310$ released cover leaves, $309$ have the first size and the one leaf with upper endpoint $1$ has the second. The singleton of Proposition~\ref{prop:appb-obstruction} has the first size. At $u=2/3$ the cap equals $1$: the five scalar $\rho_2$-upper rows disappear and the $\rho_2$ band product is stored on its own, giving $509$ inequalities if the upper endpoint is below $1$ and $503$ if it equals $1$. With $\ell_2=1/10$ and that singleton's other settings, the builder gives $519$ inequalities and $88$ PSD blocks. The number of clusters is free: there is no phase grid and no rounding.

\subsection{A feasible point obstructing this relaxation}

\begin{proposition}\label{prop:appb-obstruction}
Let $u_*=22906937/32000000$ and $d_0=29/20000$.
The rational vector specified by \texttt{release\_0717/primal\_peak.json}
is feasible for the frozen \texttt{coupled-fourier-v2} relaxation with
$(R,H,D,A,L)=(4,5,4,7,8)$, symmetric edges, default conditioning modes
$-4,\ldots,4$, no sample modes, and unconditional input $\rho_2$ band
$[0,1]$, at threshold $u_*$ on every closed $D_4$ band
$[a,b]\subset(0,1]$ containing $d_0$.
It remains feasible at every threshold $1/2<u\le u_*$.
Consequently no infeasibility certificate for this relaxation can
cover $d_0$ at such a threshold, irrespective of subdivision into
$D_4$ bands.
\end{proposition}

\begin{proof}
The point is the $238$ rational strings in \texttt{moments}, in the order fixed by \texttt{variables}. The \texttt{spec} records the parameters above and the singleton band $[d_0,d_0]$. This is an exact finite specification, not rounded decimal data. The rows are those of \texttt{release\_0717/far/fourier2.py}. Field \texttt{primary} of \texttt{EXACT\_PEAK.json} records exact feasibility on that singleton. The replay \texttt{extras/appb\_checks.py} rebuilds every row, checks the coordinate order, evaluates every scalar row and every equality over $\mathbb{Q}$, and tests each symmetric PSD block by rational Schur complements, accepting a zero pivot only when the remainder of its column is zero. It checks the box bounds, so the point lies in $[-1,1]^{238}$.

Band persistence is checked from the following identities, not inferred from the singleton. Put $z_0=1-d_0=19971/20000$. At the point, $X(4,-4)=z_0$, $X(4,-4,4,-4)=z_0^2$, and $X(j,-j,4,-4)=z_0\,X(j,-j)$ for $0\le j\le8$. The matrix $T_{ij}=X(j-i,i-j)$, $0\le i,j\le8$, is PSD, and $X(j,-j)\ge0$ for $0\le j\le4$. Reflection identifies negative frequencies with positive ones, so these values determine every localizer entry. On a band $[a,b]\ni d_0$ the two $D_4$ localizers are $(d_0-a)T$ and $(b-d_0)T$, the scalar exclusion rows carry the same nonnegative factors times $X(j,-j)$, and the exclusion band product equals $(d_0-a)(b-d_0)\ge0$. Every other row is independent of $[a,b]$. If $b=1$ the lower rows are omitted, so the point stays feasible on every enclosing band.

In the builder set $\alpha_u=1-u$, $\eta_u=3/4-u$, $c_a(u)=2-12\alpha_u^2$ and $c_b(u)=8\eta_u+16\eta_u^2$. As $u$ decreases, $\alpha_u^2$ and $c_b(u)$ increase and $c_a(u)$ decreases, so the ratio $c_b(u)/c_a(u)$ increases while $c_a(u)>0$. The cap $y_u$ of Appendix~\ref{app:ledger} is therefore nondecreasing as $u$ decreases: it equals $1$ if $c_a(u)\le0$, and otherwise $\min(1,v/1000)$ with $v$ the least nonnegative integer satisfying $v^2/10^6\ge c_b(u)/c_a(u)$. Write $T^{(2)}_{ij}=X(j-i,i-j,2,-2)$ and $\delta_\alpha=\alpha_u^2-\alpha_{u_*}^2\ge0$. The replay checks $4T+3T^{(2)}\succeq0$, checks $X(j,-j,2,-2)\ge0$ for $0\le j\le4$, and checks $X(1,-1),X(2,-2)\ge0$.

The local order entry is $c_b(u)\,T-2T^{(1)}-c_a(u)\,T^{(2)}$, where $T^{(1)}_{ij}=X(j-i,i-j,1,-1)$ does not depend on $u$. From $c_b=16\alpha^2-1$ and $c_a=2-12\alpha^2$,
\[
(c_b(u)-c_b(u_*))T-(c_a(u)-c_a(u_*))T^{(2)}=4\delta_\alpha(4T+3T^{(2)}),
\]
using $c_b(u)-c_b(u_*)=16\delta_\alpha$ and $c_a(u)-c_a(u_*)=-12\delta_\alpha$. The difference is PSD. Degree blocks gain $(u_*-u)$ times their phase blocks, which are independent of $u$ and PSD. The local $\rho_1$ and $\rho_2$-upper blocks gain $\delta_\alpha\,T$ and $(y_u^2-y_{u_*}^2)T$. Scalar rows gain the same nonnegative factors times the checked moments. The squared-order row gains $16\delta_\alpha+8(y_u-y_{u_*})X(1,-1)$. The $\rho_2$ band-product expression gains $(y_u^2-y_{u_*}^2)X(2,-2)$. At $\ell_2=0$ this expression equals the would-be scalar $\rho_2$-upper row at $j=2$. When $y_u<1$ that scalar row is emitted and the product is the same stored row; when $y_u=1$ the scalar $\rho_2$-upper family is suppressed and the product is stored under its own label. The stored value still increases by this nonnegative increment, including at $u=2/3$, where the cap equals $1$. Every remaining row is unchanged, so every threshold $1/2<u\le u_*$ is feasible.

At a feasible point the dual combination below is nonnegative. A certificate makes it strictly negative on $[-1,1]^{238}$, so none exists.
\end{proof}

This obstruction concerns only the specified finite relaxation with unconditional $\rho_2$ input. The vector is not asserted to arise from a graph, a macrostate, or a phase measure, and it is not a counterexample to synchronization. It does not rule out a larger moment system, an extra conditioning, or a stronger analytic near-region estimate that reaches $d_0$. In particular $u_*>0.71584$, so no certificate at a threshold $u\le0.71584$ can cover $d_0$ inside this relaxation. That is not a barrier for every extension of the method.

\subsection{Dual certificates}

Write the relaxation as $g_j(v)\ge0$, $h_i(v)=0$ and $H_l(v)\succeq0$. A certificate gives nonnegative rationals $\lambda_j$, rationals $\kappa_i$ of either sign, and PSD matrices $Y_l$, supplied as $GG^{\mathsf T}/s^2$ or as $\sum w\,vv^{\mathsf T}/s^2$ with integer $G$ or $v$, denominator $s>0$, and $w\ge0$. The combination
\begin{equation}
\sum_j\lambda_jg_j(v)+\sum_i\kappa_ih_i(v)+\sum_l\langle Y_l,H_l(v)\rangle
=b_0+\sum_kr_kv_k
\end{equation}
is an affine identity. The left side is nonnegative at every feasible point. If $b_0+\sum_k|r_k|<0$, the right side is negative throughout $[-1,1]^n$, and no feasible point exists. The test is exact rational arithmetic. Solver output does not enter it.

Format \texttt{coupled-fourier-certificate-2} is checked by \texttt{far/verify2.py}, with $Y=GG^{\mathsf T}/s^2$. Format \texttt{weighted-fourier-certificate-4} is checked by \texttt{far/verify\_leaf4.py}, which passes a coupled leaf to \texttt{verify2.py}. Each rebuilds the rows in \texttt{far/fourier2.py} and matches the variable list. Once the spec, the variable manifest, the constraint counts, the multiplier types and signs, and the Gram labels and shapes have passed validation, a leaf is accepted if and only if $b_0+\sum_k|r_k|<0$. An ill-formed file can be rejected before that residual is computed. On format \texttt{fourier-band-cover-4}, \texttt{far/verify4.py} requires every leaf to pass at a threshold no higher than the cover and rejects an input $|\rho_2|$ band other than $[0,1]$. The routine \texttt{covers} accepts when the bands cover the requested interval by exact closed fractions, overlap allowed, with no gap and no tolerance. A band may start below the requested lower endpoint. A threshold $u'\le\mus$ suffices: $\mu(T)\ge\mus$ implies $\mu(T)\ge u'$.

\subsection{The cover}

The released cover is $310$ bands tiling $[\Delta,1]$ with no gaps (Table~\ref{tab:cover}). Every adjacent pair meets exactly, and the first band starts exactly at $\Delta$. Every band uses the cutoffs above, symmetric edge blocks, and input $0\le|\rho_2|\le1$. Four inherited bands use thresholds below $\mus$. Solvers in exact active-basis coordinates only propose candidates. Acceptance is the rational test above. The README of the verification code points to \texttt{SHA256SUMS} and to \texttt{./run\_checks.sh}.

\begin{table}[tb]
\caption{The far cover at $\mus=717/1000$. ``Weighted'' certificates give PSD duals as sums of rank-one
terms; ``coupled'' certificates give them as $GG^{\mathsf T}$. The group endpoints are
$b_1=7/138240\approx5.0637\times10^{-5}$ and $b_2=152587890625/101559956668416\approx1.50244\times10^{-3}$.}
\label{tab:cover}
\begin{ruledtabular}
\begin{tabular}{rlll}
leaves & $D_4$ range & format & threshold\\
\hline
135 & $[4.9\times10^{-6},\,b_1]$ & weighted & $0.717$\\
36 & $[b_1,\,10^{-4}]$ & weighted & $0.717$\\
105 & $[10^{-4},\,b_2]$ & weighted & $0.717$\\
30 & $[b_2,\,0.04]$ & coupled & $0.717$\\
1 & $[0.04,0.06]$ & coupled & $0.716$\\
1 & $[0.06,0.1]$ & coupled & $0.715$\\
1 & $[0.1,0.2]$ & coupled & $0.713$\\
1 & $[0.2,1]$ & coupled & $0.710$\\
\end{tabular}
\end{ruledtabular}
\end{table}

%==========================================================================================
\section{Validity of the far relaxation}\label{app:ledger}

The rows below are exactly the release call to \texttt{fourier2.py}, which fixes the exact index sets and block sizes: symmetric edges, empty \texttt{sample\_modes}, and the cutoffs of Appendix~\ref{app:far}, including Proposition~\ref{prop:appb-obstruction}. Sample-mode blocks are implemented, unused, and not derived. Fix $1/2<u<3/4$ and $0<\delta_-\le D_4\le\delta_+\le1$; the builder rejects any other specification. An optional input $\ell_2\le|\rho_2|\le h_2$ is allowed. The release passes $[0,1]$, and the cap may then lower the upper end. Let $T$ be linearly stable with $d_a\ge u$ for every cluster and with $D_4$ in the band. (The builder checks the threshold and the band endpoints of the specification; that $T$ lies in the band is a hypothesis, used to derive the constraints.) Rotations and reflections preserve the hypotheses. Moments are those of Appendix~\ref{app:far}, including $\rho_k=\int e^{ik\phi}\,\dd\pi$.

\emph{Canonical form.} These rules hold for every macrostate, so no symmetry of the input is assumed. Nonzero total frequency averages to zero. Factors $\rho_0=1$ may be dropped, and $X()=1$. The $\rho$ indices may be permuted, the edge indices exchanged, and all indices negated; the stored key is the lexicographically smaller representative. $N(0,0;)$ is the mass of $\nu$, not one.

\begin{lemma}[Gram matrices]\label{lem:gram}
Gram matrices of a nonnegative measure are PSD, as is $\mathbb{E}[\overline{\xi_i}\xi_j]$ for complex numbers $\xi_i$ computed from a macrostate. Nonnegative multiples, principal sub-blocks, and averages of PSD matrices are PSD. Deleting a zero row, a duplicate scalar row, or a conjugate block only weakens the constraint. Suppose, in addition, that the average includes a uniform rotation and an independent fair reflection (as in Appendix~\ref{app:far}), that each feature is homogeneous of one total frequency, and that reflection, applied to the measure or form together with the macrostate, sends the Gram matrix to its entrywise conjugate. Then the averaged matrix splits into real symmetric blocks, one total frequency each.
\end{lemma}

\begin{proof}
For a nonnegative measure $\mu$ and complex $f_i$, the quadratic form is $\int|\sum_iv_if_i|^2\,\dd\mu$. For macrostate values $\xi_i$ it is $\mathbb{E}|\sum_iv_i\xi_i|^2$: the average of the copies' Gram matrices, not the Gram matrix of the averaged moments. The conjugate of a PSD block is PSD, so omitting it weakens the constraint. Under the additional hypotheses, rotation by $\theta$ multiplies total frequency $t$ by $e^{it\theta}$, so a cross term between different totals averages to zero. The average also includes an independent fair reflection, which replaces each block by its entrywise conjugate, so the average of the two is real symmetric. These hypotheses are sufficient, not necessary. Without the conjugation condition reality can fail: the constant macrostate values $1$ and $i$ give a PSD Gram matrix whose imaginary off-diagonal entries are unchanged by reflection.
\end{proof}

Table~\ref{tab:families} lists the entries. Except the Hessian, each row is Lemma~\ref{lem:gram} for the input shown, with $\rho_r$ included before the average. The splitting and reality hypotheses hold for these Fourier features, including both parity projections: each has real coefficients, and the measure or form is transformed with the macrostate. The release uses the parity rows. Conditioning indices lie in a symmetric set $S\ni0$ inside $\{-R,\ldots,R\}$; the release uses every such mode. Mass indices lie in $S$, and only totals $t\ge0$ are kept. Edge endpoints lie in $0,\ldots,A$. Phase, degree, and diagonal tests lie in $[-D,D]$, and Hessian tests lie in $[-H,H]\setminus\{0\}$; those four families also keep only totals $t\ge0$. Symmetric edge and complement use $0\le a,b\le A$ and $r\in S$ with total $t=a+b+r\le A$, and they do not discard negative totals. Without symmetrization the same endpoints and modes are used and every total is considered before scalar rows are deduplicated; the release does not emit those blocks. The exact index sets and block sizes are those of \texttt{far/fourier2.py}.

\begin{table*}[t]
\caption{Gram entries. In a parity row, $\varepsilon=\pm1$ and the second summand swaps the test frequencies. Hessian: text.}
\label{tab:families}
\begin{ruledtabular}
\scriptsize
\begin{tabular}{llp{0.46\textwidth}}
family & input & entry\\
\hline
mass $t\ge0$ & $\rho_a\rho_b$, $a\le b$ & $X(-a,-b,c,d)$\\
edge & $\nu$; $e^{i(a\phi+b\psi)}\rho_r$ & $N(c-a,d-b;-r,s)$\\
complement & $\pi\otimes\pi-\nu$ & $X(c-a,d-b,-r,s)-N(c-a,d-b;-r,s)$\\
sym.\ edge $(t,\varepsilon)$ & $(f+\varepsilon f_{\mathrm{swap}})/2$, $t\le A$ & $\bigl(N(c-a,d-b;-r,s)+\varepsilon N(d-a,c-b;-r,s)\bigr)/2$\\
sym.\ complement & same features on $\pi\otimes\pi-\nu$ & preceding formula, $X-N$ for $N$\\
degree $t\ge0$ & $(d-u)\,\dd\pi$; $e^{ik\phi}\rho_r$ & $N(l-k,0;-r,s)-u\,X(l-k,-r,s)$\\
diagonal $t\ge0$ & $c\,\dd\pi$; same & $\tfrac12\sum_{v=\pm1}N(l-k+v,-v;-r,s)$\\
phase $t\ge0$ & $\dd\pi$; same & $X(l-k,-r,s)$\\
Hessian $t\ge0$ & $Q$; $0<|k|\le H$ & $\tfrac12\sum_{v=\pm1}\bigl(N(l-k+v,-v;-r,s)-N(-k+v,l-v;-r,s)\bigr)$\\
\end{tabular}
\end{ruledtabular}
\end{table*}

\paragraph*{Symmetrized edges.}
Let $f_{abr}=e^{i(a\phi+b\psi)}\rho_r$. The Gram integral of $(f_{abr}+\varepsilon f_{bar})/2$ expands into four monomials divided by four. Exchange symmetry of $\nu$ and of $\pi\otimes\pi$ pairs outer monomials and pairs middle monomials, leaving the factor $1/2$ and no factor $\sqrt2$. Even and odd features are orthogonal. An odd feature with equal edge frequencies vanishes, so the builder keeps $a<b$, or $a=b$ with $\varepsilon=+1$, and drops totals above $A$. Retained blocks are PSD because $0\le P_{ab}\le1$.

\paragraph*{Degree, diagonal, and Hessian.}
With $f_{kr}(\phi)=e^{ik\phi}\rho_r$, the degree, diagonal, and phase entries are Gram matrices of weights $x_a(d_a-u)$, $x_ac_a$, and $x_a$. Use $d_a\ge u$ and $c_a\ge0$ on each cluster, with no lower bound on $x_a$. Then $d_a=\sum_bx_bP_{ab}$ is the degree entry, and $\cos t=(e^{it}+e^{-it})/2$ gives Fourier coefficient $\tfrac12(n_{m+1,-1}+n_{m-1,1})$ for $c\,\dd\pi$.

Omit $k=0$, since $Q$ annihilates constants. For $0<|k|\le H$ the pre-average entry is
\[
\tfrac12\int\cos(\phi-\psi)\,\overline{(f_{kr}(\phi)-f_{kr}(\psi))}\,(f_{ls}(\phi)-f_{ls}(\psi))\,\dd\nu.
\]
Without $\tfrac12$ this is the ordered-pair form $Q$ in Eq.~\eqref{eq:Q}, so the displayed matrix is PSD. A real PSD form stays PSD on complex vectors, real and imaginary parts separating. On one macrostate $\rho_r$ is a fixed number, so $e^{ik\varphi}\rho_r$ is one vector in $Q\ge0$. Eight monomials arise from the cosine and the two differences; $n_{pq}=n_{qp}$ pairs them. Two factors $\tfrac12$ yield the table coefficient, a pair doubling one monomial and $\tfrac14\cdot2=\tfrac12$. Atoms $\phi=\psi$ contribute $0$. The three tests $e^{i\varphi}$, $\rho_4e^{-3i\varphi}$ and $\rho_{-4}e^{5i\varphi}$ have total frequency $1$, with $(k,r)=(1,0)$, $(-3,4)$ and $(5,-4)$, so one Hessian block tests them simultaneously.

\paragraph*{Equilibrium.}
For $r,s\in S$ and $k=-r-s$, the row $N(k+1,-1;r,s)-N(k-1,1;r,s)=0$ is, before averaging,
\[
2i\,\rho_r\rho_s\sum_ax_ae^{ik\varphi_a}F_a=0,
\]
by $\sin\theta=(e^{i\theta}-e^{-i\theta})/(2i)$ and $F_a=\sum_bx_bP_{ab}\sin(\varphi_a-\varphi_b)$. Only $F=0$ is used, and $|k|\le2R$.

\subsection{The semicircle bound, the order bounds, and the cap}

Set $a_0=1-u$, $\eta=3/4-u$, $p=|\rho_1|^2$, $y=|\rho_2|$, $q=y^2$,
$A_0=2-12a_0^2=\tfrac54-6\eta-12\eta^2$ and $B_0=8\eta+16\eta^2=16a_0^2-1$.
These match Lemma~\ref{lem:order} because $a_0=\eta+\tfrac14$.

\paragraph*{Semicircle: $p\le a_0^2$.}
Rotate so that $\rho_1\ge0$. Equilibrium gives $\rho_1-c_ae^{i\varphi_a}=\sum_bx_b(1-P_{ab})e^{i\varphi_b}$, of modulus at most $1-d_a\le a_0$. If $\rho_1>a_0$, then $c_a\cos\varphi_a\ge\rho_1-a_0>0$, so $c_a\ge0$ forces every phase into $(-\pi/2,\pi/2)$. Edges $P_{ab}>0$ form one component, because the component of $a$ has mass at least $d_a\ge u>1/2$. At a maximal phase every difference lies in $[0,\pi)$, so every sine in $F_a$ is nonnegative and the sine toward a strictly smaller phase is positive. Thus the maximal phase has no edge out. Connectivity forces synchronization, hence $D_4=0$, contradicting $\delta_->0$. So $p\le a_0^2$. No spectral gap is used.

\paragraph*{Quartic bound.}
Let $Q_{ab}=1-P_{ab}$, $q_a=\sum_bx_bQ_{ab}\le a_0$, $t_{ab}=\cos(\varphi_a-\varphi_b)$ and $w_b=x_bQ_{ab}\ge0$, possibly fractional. Tests of $Q$ on $\cos\varphi$ and $\sin\varphi$ give $\sum_{a,b}N_{ab}(t_{ab}-t_{ab}^2)\ge0$, since the two squared increments sum to $2(1-t_{ab})$. Since $N_{ab}=x_ax_bP_{ab}$, replacing $P$ by $1$ uses the complete-pair weights $x_ax_b$ and produces $p-(1+q)/2$, because $\sum_{a,b}x_ax_bt_{ab}=p$ and $\cos^2\theta=(1+\cos2\theta)/2$ turns $\sum_{a,b}x_ax_bt_{ab}^2$ into $(1+q)/2$. Subtracting the nonnegative edge sum yields
\begin{equation}
\sum_{a,b}x_ax_bQ_{ab}(t_{ab}-t_{ab}^2)\le p-\frac{1+q}{2}.
\label{eq:O1}
\end{equation}
In the same gauge, $\sum_bw_b\sin(\varphi_a-\varphi_b)=\rho_1\sin\varphi_a$. Cauchy--Schwarz twice, the split $\sum_bw_bt_{ab}^2=q_a-\sum_bw_b\sin^2$, and $q_a\sum_bw_b\sin^2\ge p\sin^2\varphi_a$ give
\begin{equation}
\begin{aligned}
\Bigl(\sum_bw_b|t_{ab}|\Bigr)^2
&\le q_a\sum_bw_bt_{ab}^2
\le q_a^2-p\sin^2\varphi_a\\
&\le a_0^2-p\sin^2\varphi_a,
\end{aligned}
\label{eq:cs}
\end{equation}
where $0\le q_a\le a_0<1/2$ supplies $q_a^2\le a_0^2$. If $q_a=0$ there is no division. On $[-1,1]$, $t-t^2+2|t|$ equals $t(3-t)$ for $t\ge0$ and $-t(1+t)$ for $t<0$, hence $t-t^2\ge-2|t|$. The left side of Eq.~\eqref{eq:O1} is at least $-2\sum_ax_aS_a$ for $S_a=\sum_bw_b|t_{ab}|\le\sqrt{a_0^2-p\sin^2\varphi_a}$. Concavity of the square root and $\sum_ax_a\sin^2\varphi_a=(1-\mathrm{Re}\,\rho_2)/2\ge(1-y)/2$ give $\sum_ax_aS_a\le\sqrt{a_0^2-p(1-y)/2}$. Therefore
\[
1-2p+q\le4\sqrt{a_0^2-p(1-y)/2}=\sqrt{16a_0^2-8p(1-y)}.
\]
The left side exceeds $1/2$ because $p\le a_0^2<1/4$. Squaring gives
\begin{equation}
(1-2p+q)^2+8p(1-y)\le16a_0^2.
\label{eq:O2}
\end{equation}

\paragraph*{Linear bound.}
Expanding Eq.~\eqref{eq:O2} produces $E\ge0$, where $E=B_0-4p+8py-2q+4pq-4p^2-q^2$. From $8y\le2+8y^2$, whose slack is $8(y-\tfrac12)^2$, one has $8py\le2p+8pq$, so $E\le E_1$ with $E_1=B_0-2p-2q+12pq-4p^2-q^2$. Then $-4p^2-q^2\le0$ and $12pq\le12a_0^2q$, the second because $0\le p\le a_0^2$, yield $E_1\le B_0-2p-A_0q$. Thus $0\le E\le E_1\le B_0-2p-A_0q$, that is $2p+A_0q\le B_0$, without assuming $A_0>0$.

\paragraph*{The cap.}
If $A_0>0$, then $q\le B_0/A_0$. Let $v$ be the least nonnegative integer with $v^2/10^6\ge B_0/A_0$: the integer square root of $\lfloor 10^6B_0/A_0\rfloor$, plus one when that square is strictly too small. Exact squares are included. Set $h_2'=\min(h_2,v/1000)$; a quotient above $1$ does not raise $h_2\le1$. If $A_0\le0$, the input upper bound is unchanged. Since $p\ge0$ and $y\le h_2'$, substituting $h_2'$ for $y$ weakens Eq.~\eqref{eq:O2} to $16a_0^2-(1-2p+q)^2-8(1-h_2')p\ge0$. At $u=717/1000$ and unconditional input, $h_2'=521/1000$. With an optional input bound the same threshold gives $h_2'=\min(h_2,521/1000)$.

\subsection{Localizers, band products, and boxes}

Let $z=|\rho_4|^2$. The scalars $1-\delta_--z$, $z-(1-\delta_+)$, $a_0^2-p$, $B_0-2p-A_0q$, $h_2'^2-q$, and $q-\ell_2^2$ are nonnegative. For $j=0,\ldots,R$, multiply each emitted scalar by $|\rho_j|^2$. Omit the scalar for the upper bound on $D_4$ (equivalently the lower bound $z\ge1-\delta_+$) if $\delta_+=1$, the scalar $\rho_2$-upper row if $h_2'=1$, and the lower $\rho_2$ row if $\ell_2=0$. The release cap satisfies $h_2'<1$. For $L\ge1$ the emitted scalars multiply $(|\rho_{j-i}|^2)_{0\le i,j\le L}$, the Gram matrix of $e^{i(i\phi+(L-i)\psi)}$ in $L^2(\pi\otimes\pi)$, hence PSD by Lemma~\ref{lem:gram}. The localizer for the upper bound on $D_4$ is skipped when $\delta_+=1$, and the lower $\rho_2$ localizer when $\ell_2=0$. The scalar $b+\sum_kw_k|\rho_k|^2$ has entries $b\,X(j-i,i-j)+\sum_kw_kX(j-i,i-j,k,-k)$. The local $\rho_2$-upper block is emitted even if $h_2'=1$.

On the band, $1-\delta_+\le z\le1-\delta_-$ and $\ell_2^2\le q\le h_2'^2$, so both factors of each product are nonnegative before averaging, including when a stored row duplicates another. Thus $\mathbb{E}[(z-1+\delta_+)(1-\delta_--z)]\ge0$ and $\mathbb{E}[(q-\ell_2^2)(h_2'^2-q)]\ge0$. The two expectations equal
\begin{align*}
&(2-\delta_--\delta_+)X(4,-4)-X(4,-4,4,-4)\\
&\qquad-(1-\delta_-)(1-\delta_+),\\
&(\ell_2^2+h_2'^2)X(2,-2)-X(2,-2,2,-2)-\ell_2^2h_2'^2.
\end{align*}
If $\delta_+=1$, the first equals the scalar exclusion row at $j=4$ and is stored once. If $\ell_2=0$ and $h_2'<1$, the second equals the scalar $\rho_2$-upper row at $j=2$ and is stored once. Only the last cover leaf has $\delta_+=1$. Every cover leaf has $\ell_2=0$ and $h_2'<1$. The quartic row is the weakened Eq.~\eqref{eq:O2}, expanded before averaging:
\begin{align*}
16a_0^2&-\bigl(1-4X(1,-1)+2X(2,-2)+4X(1,-1,1,-1)\\
&-4X(1,-1,2,-2)+X(2,-2,2,-2)\bigr)\\
&-8(1-h_2')X(1,-1)\ge0.
\end{align*}
The coordinate $X(1,-1,1,-1)$ is the fourth moment $\mathbb{E}[p^2]$. On one macrostate $p$ is invariant under rotation and reflection, so $\mathbb{E}[p^2]=(\mathbb{E}p)^2$. The relaxation keeps this coordinate and does not enforce equality with $X(1,-1)^2$. The rows $1\pm v\ge0$ hold because $|\prod_k\rho_k|\le1$ and $|n_{kl}|$ is at most the mass of $\nu$. The average and the real part do not increase the modulus, and equilibrium variables are included.

Each emitted family is Lemma~\ref{lem:gram}, a Gram matrix of $Q$, the expectation of a scalar inequality above, or $F=0$. Nonsynchronization enters only through $\delta_->0$ in the semicircle bound. No individual mass has a positive lower bound, and the derivation uses no strict spectral gap, no phase grid, and no equal-mass hypothesis.

\section{Proof of Theorem~\ref{thm:shell}}\label{app:shell}

Assume $T$ is linearly stable with $\mu(T)\ge\mus$ and $0<D_4\le\Delta$.
The exact inputs are in Table~\ref{tab:constants} and \texttt{shell\_717.json}.
The released \texttt{check\_final.py} verifies the scalar tests and cover below with exact
\texttt{Fraction} arithmetic, and its \texttt{algebra(k)} verifies the trial polynomials indicated below.
Tables~\ref{tab:derived} and~\ref{tab:margins} report derived values and margins. Table~\ref{tab:shell-chain} locates the scalar tests.

\begin{figure}[t]
\centering
\includegraphics[width=\columnwidth]{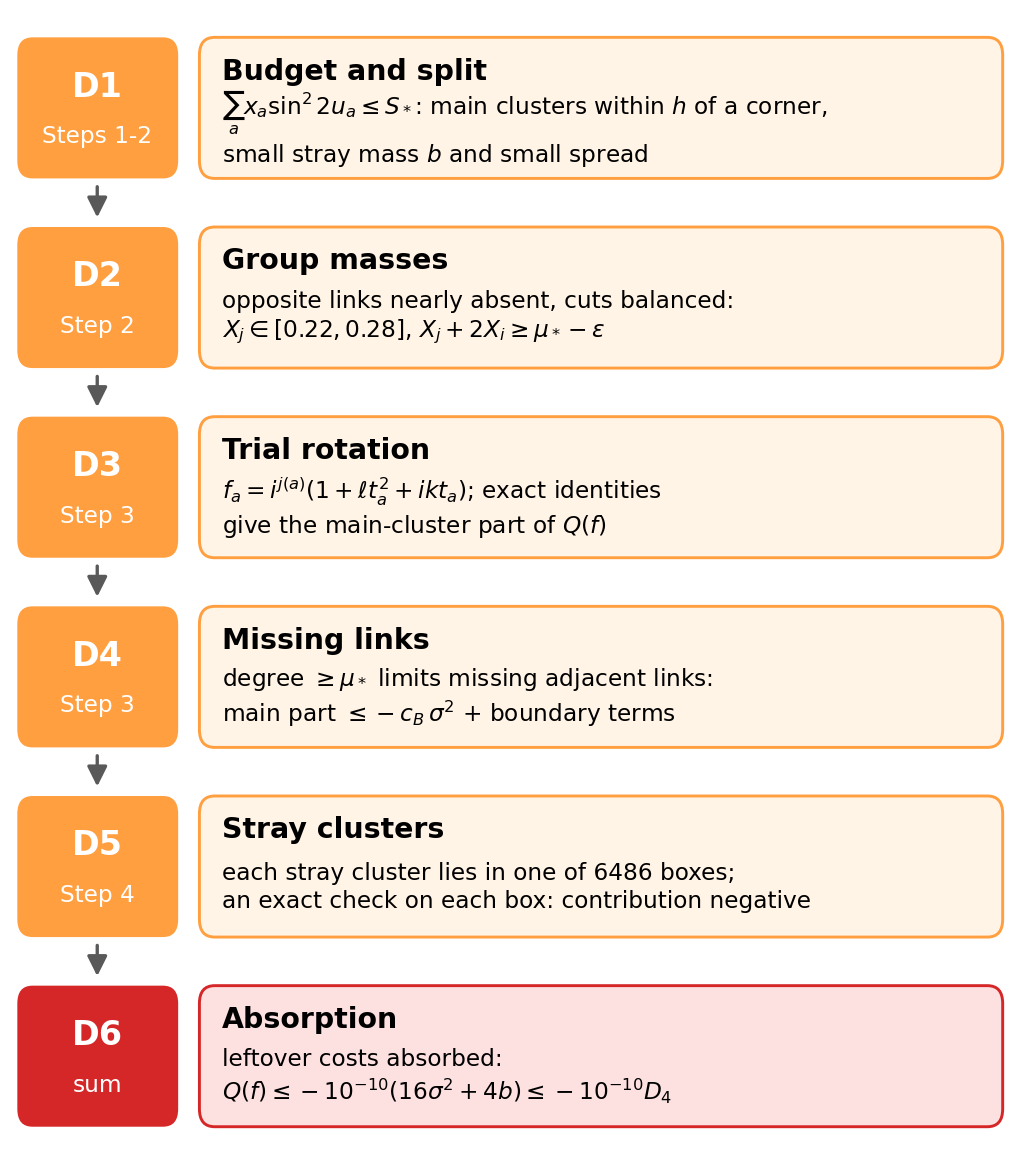}
\caption{Roadmap of the near-region proof. Each part of this Appendix (left) proves one step of
Sec.~\ref{sec:shell}; the last combines them into Eq.~\eqref{eq:shellQ}.}
\label{fig:flow}
\end{figure}

\begin{figure}[t]
\centering
\includegraphics[width=\columnwidth]{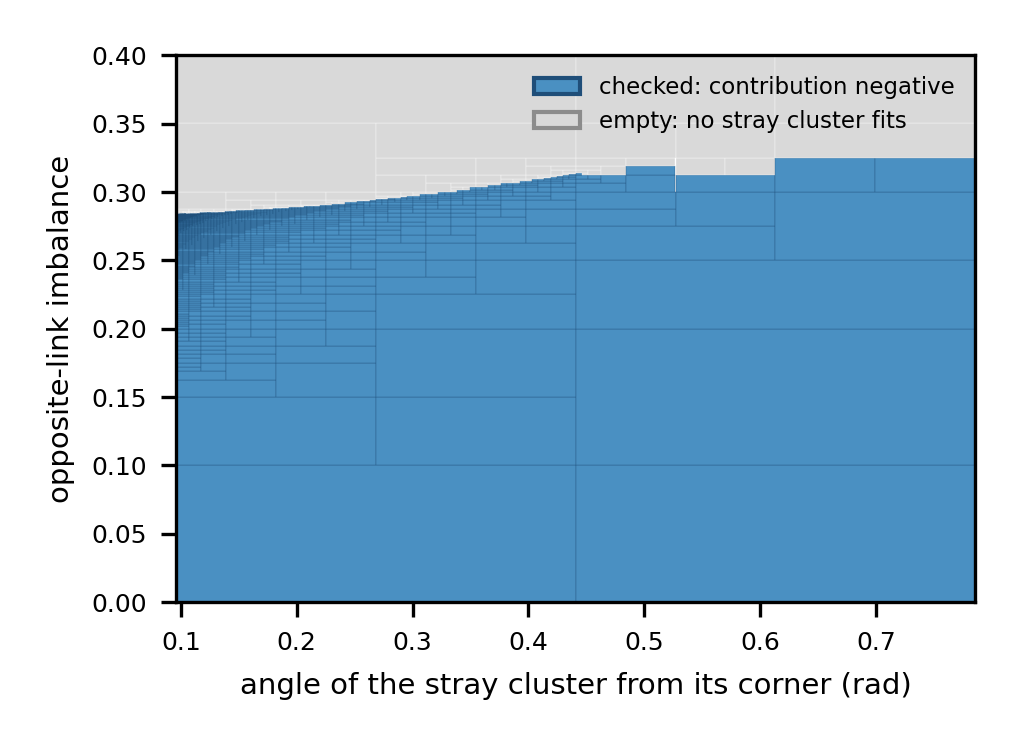}
\caption{The stray-cluster cover (Subsection~\ref{app:D5}). After folding by quarter turns and a reflection, a stray
cluster is described by two numbers: its angle from the nearest corner and the imbalance of its links to
opposite groups. The rectangle of possible values is split into 6486 boxes, refined where the bound is
tight. On 6267 boxes (blue) an exact check shows that the idealized contribution to $Q(f)$ is negative
throughout; on the other 219 (gray) no stray cluster can exist.}
\label{fig:boxes}
\end{figure}

\paragraph*{Roadmap.} Figure~\ref{fig:flow} shows how the six parts below follow Steps~1--4 of
Sec.~\ref{sec:shell}. Subsections~\ref{app:D1}--\ref{app:D2} pin down the shape of a linearly stable macrostate in the near
region: a small stray mass, a small spread, and four nearly equal groups with almost no opposite links
(Steps~1--2). Subsections~\ref{app:D3}--\ref{app:D4} bound the main clusters' contribution to $Q(f)$ for the trial rotation
(Step~3). Subsection~\ref{app:D5} handles the stray clusters with the box cover of Fig.~\ref{fig:boxes} (Step~4),
and Subsection~\ref{app:absorb} adds everything up. Each step is elementary (Taylor bounds, the Cauchy--Schwarz and
Young inequalities); the length comes from keeping every constant explicit.

\begin{table}[tb]
\caption{Parameters of the near-region proof at $\mus=717/1000$. Each is an exact rational; decimals are
for orientation.}
\label{tab:constants}
\begin{ruledtabular}
\begin{tabular}{lll}
symbol & meaning & value\\
\hline
$\mus$ & target degree & $717/1000$\\
$\Delta$ & near-region cutoff & $49/10^7$\\
$h$ & main-cluster cutoff & $483/5000$\\
$\alpha$ & mean/$L^1$ bound & $557004883/10^{12}$\\
$\beta$ & stray mass & $33246759/10^{12}$\\
$\Phi$ & main-cluster second moment & $310244123969/10^{18}$\\
$m_0$ & coarse group floor & $21138230883/10^{12}$\\
$M$ & group cap & $283094754719/10^{12}$\\
$q$ & stray degree & $716131960343/10^{12}$\\
$\gamma$ & stray penalty & $11/1250000$\\
$k$ & slope & $1377/1000$\\
$\tau$ & Young shift & $71/500000$\\
$g_B$ & main-cluster floor & $949689053/10^{10}$\\
$d$ & $D_4$ coefficient & $1/10^{10}$\\
\end{tabular}
\end{ruledtabular}
\end{table}

\subsection{Gauge and coarse group masses}\label{app:D1}

Rotate to $\rho_4=r\ge0$ and choose the nearest corner,
$\varphi_a=j(a)\pi/2+u_a$, $|u_a|\le\pi/4$.
Since $\sin^2(2u_a)=(1-\cos4u_a)/2$, Eq.~\eqref{eq:S1} bounds its mass average by $S_*$.
Use $\B_j=\{a:j(a)=j,\ |u_a|\le h\}$, $\B=\bigcup_j\B_j$, $\E=\B^c$,
$b=x(\E)$ and $X_j=x(\B_j)$. With
$s_7(z)=z-z^3/6+z^5/120-z^7/5040\le\sin z$ for $0\le z\le\pi/2$ (the series terms decrease in magnitude there),
monotonicity of $\sin2v$ and $v/\sin2v$ on $[0,\pi/4]$ gives
\begin{align}
 b&\le S_*/s_7(2h)^2\le\beta,\label{eq:A-S2}\\
 \sum_\B x_au_a^2&\le h^2S_*/s_7(2h)^2\le\Phi.\label{eq:A-S3}
\end{align}
Set $\bar u=\sum_\B x_au_a/(1-b)$, $t_a=u_a-\bar u$, $V=\sum_\B x_at_a^2$,
$T_h=h+\alpha$, and $D=2T_h$. Centering and Cauchy--Schwarz using $\Phi\le\alpha^2(1-\beta)$ give
\begin{align*}
 \sum_\B x_at_a&=0,\quad V\le\Phi,\quad |t_a|\le T_h,\\
 |\bar u|&\le\sqrt{\Phi/(1-\beta)}\le\alpha,\\
 U_t&=\sum_\B x_a|t_a|\le\sqrt V,\\
 U_\B&=\sum_\B x_a|u_a|\le\sqrt\Phi\le\alpha.
\end{align*}
Let $\widehat X_l=\sum_j X_ji^{lj}$.
Stray clusters cost at most $b$ and main phase shifts at most $lU_\B$, so
$|\rho_l-\widehat X_l|\le b+lU_\B$.
Using $\widehat X_0=1-b$ and $\widehat X_3=\overline{\widehat X_1}$, Fourier inversion gives
\[
 X_j\ge\tfrac14-|\rho_1|/2-|\rho_2|/4-b-U_\B.
\]
At $\eta=3/4-\mus=33/1000$, Cauchy--Schwarz in Eq.~\eqref{eq:order} yields
\begin{equation}
 X_j\ge\tfrac14-\sqrt{B_0(1/8+1/(16A_0))}-\alpha-\beta\ge m_0.
 \label{eq:shell-floor}
\end{equation}
The coarse-floor test uses $A_0>0$ and $m_0>0$, so all four main groups are occupied.

\begin{table}[t]
\caption{Derived quantities in the near-region proof (decimal displays of exact rationals).}
\label{tab:derived}
\begin{ruledtabular}
\begin{tabular}{ll}
quantity & value\\
\hline
$S_*$ & $49/39980000\approx1.2256\times10^{-6}$\\
floors $m_0$, $m_1$ & $0.02114$, $0.21590$\\
floors $m_2$, $m_3$ & $0.21686$, $0.21686$\\
$O_*$, $\varepsilon$ & $2.055\times10^{-5}$, $1.803\times10^{-4}$\\
$c'$ & $0.716820$\\
$L$, $L-\mathcal{A}M$ & $3.8536$, $0.08833$\\
$F_*(s)$ at endpoints & $0.4415$, $0.05211$, $7.63\times10^{-8}$\\
$R_w$ & $71.617$\\
opposite coefficient & $0.0017966$\\
$c_B$, $K$ & $0.030532$, $29.021$\\
$C_{\min}$ & $3377/250$\\
$B'$, $E'$ & $0.029567$, $1.0562\times10^{-5}$\\
$4B'E'-S_*$ & $2.347\times10^{-8}$\\
\end{tabular}
\end{ruledtabular}
\end{table}

\subsection{Opposite mass and group balance}\label{app:D2}

All group indices are modulo four. In sums over main clusters $\sum_{\rm adj}$ counts each adjacent pair once, oriented from $\B_j$ to $\B_{j+1}$, including the wraparound. The sums $\sum_{\rm same}$ and $\sum_{\rm opp}$ are ordered, and $a<b$ means unordered in a fixed ordering. Diagonal pair differences vanish. All bounds use $0\le N_{ab}\le x_ax_b$.
Put $d_{ab}=u_a-u_b=t_a-t_b$. Write $K_j,Y_j$ for the sums of $N_{ab},N_{ab}\sin d_{ab}$ on the $j$ to $j+1$ cut, and $O_j,O_{c,j}$ for the sums of $N_{ab},N_{ab}\cos d_{ab}$ on the $j$ to $j+2$ cut. Then $O_j=O_{j+2}$, $O=O_0+O_1$, $O_c=O_{c,0}+O_{c,1}$, $Y=\sum_jY_j$, and the ordered opposite mass is $O_o=2O$.

\paragraph*{Opposite mass.}
The group indicator gives
\begin{multline*}
 0\le\tfrac12 Q(1_{\B_j})=Y_j+Y_{j-1}-O_{c,j}\\
 +\sum_{a\in\B_j,e\in\E}N_{ae}\cos(\varphi_a-\varphi_e).
\end{multline*}
The stray sum over $j$ is at most $(1-b)b\le b$, so $2O_c\le2Y+b$.
Summing $x_au_aF_a=0$ on the main clusters and pairing by the stated orientations gives
\begin{multline*}
 \sum_{\rm adj}N_{ab}d_{ab}\cos d_{ab}
 =\sum_{\substack{a<b\\\rm same}}N_{ab}d_{ab}\sin d_{ab}\\
 -\sum_{\substack{a<b\\\rm opp}}N_{ab}d_{ab}\sin d_{ab}
 +\sum_{a\in\B,e\in\E}N_{ae}u_a\sin(\varphi_a-\varphi_e).
\end{multline*}
Here $|d_{ab}|\le2h<1$ and $0\le d_{ab}\sin d_{ab}\le d_{ab}^2$, so the opposite term can be dropped and the boundary costs at most $hb$. The unordered moment identity is
\begin{multline}
 \sum_{a<b\in\B}x_ax_bd_{ab}^2\\
 =(1-b)\sum_\B x_au_a^2-\Bigl(\sum_\B x_au_a\Bigr)^2\le\Phi.
 \label{eq:shell-pair-second}
\end{multline}
Writing $Y=\sum_{\rm adj}N_{ab}[d_{ab}\cos d_{ab}+(\sin d_{ab}-d_{ab}\cos d_{ab})]$, use
$|\sin v-v\cos v|\le |v|^3/3$ by integration of $v\sin v$.
The same and adjacent contributions are bounded by their unordered second moments with coefficients $1$ and $2h/3$, so
\begin{align*}
 Y&\le(1+2h/3)\Phi+hb,\\
 O&\le O_*=\frac{(1+2h/3)\Phi+(h+1/2)\beta}{1-2h^2}.
\end{align*}
The second bound uses $O_c\ge(1-2h^2)O$ and the positive denominator test. This precedes the cap and bootstrap.

\paragraph*{Cap and cut balance.}
For $a\in\B_j$ let $o_a=\sum_{b\in\B_{j+2}}x_bP_{ab}$.
Averaging $d_a\le1-X_{j+2}+o_a$ gives $\mus X_j\le X_j(1-X_{j+2})+O_j$.
For any established floor $m'>0$, this gives $X_{j+2}\le1-\mus+O_*/m'$.
Summing $x_aF_a=0$ on $\B_j$ cancels internal terms. The sine factors on the forward, backward and opposite cuts are $-\cos d_{ab},+\cos d_{ab},-\sin d_{ab}$. Hence
\begin{align*}
 K_j-K_{j-1}
 &=\sum_{\B_j\times\B_{j+1}}N_{ab}(1-\cos d_{ab})\\
 &\quad-\sum_{\B_j\times\B_{j-1}}N_{ab}(1-\cos d_{ab})\\
 &\quad-\sum_{\B_j\times\B_{j+2}}N_{ab}\sin d_{ab}\\
 &\quad+\sum_{a\in\B_j,e\in\E}N_{ae}\sin(\varphi_a-\varphi_e).
\end{align*}
The first two cuts are disjoint unordered pair sets, so their absolute costs total at most $\Phi/2$ by Eq.~\eqref{eq:shell-pair-second} and $1-\cos v\le v^2/2$.
The remaining costs give $|K_j-K_{j-1}|\le\Phi/2+2hO_j+X_jb$.
For either neighbor $i$, its cut has mass at most $X_jX_i$ and the other at most that plus the imbalance. Thus
\begin{align*}
 \mus X_j&\le X_j^2+K_j+K_{j-1}+O_j+X_jb\\
 &\le X_j^2+2X_jX_i+(1+2h)O_*+\Phi/2+2X_jb.
\end{align*}
Divide by $X_j\ge m'$ to obtain Eq.~\eqref{eq:S2} in the form
\[
 X_j+2X_i\ge\mus-\varepsilon(m'),\quad
 \varepsilon(m')=2\beta+\frac{(1+2h)O_*+\Phi/2}{m'}.
\]
Starting from $m_0$, apply the cap to $X_j$ and this bound to $X_i$ for $r=0,1,2$:
\[
 \begin{split}
 M_r&=1-\mus+O_*/m_r,\qquad \varepsilon_r=\varepsilon(m_r),\\
 m_{r+1}&=\max\{m_r,(\mus-\varepsilon_r-M_r)/2\}.
 \end{split}
\]
The checker repeats these three updates and tests $M\ge1-\mus+O_*/m_3$.
Hence $m_3\le X_j\le M$. Set $\varepsilon=\varepsilon(m_3)$ and $c'=\mus-\varepsilon$.

\subsection{Trial identities and the main-cluster estimate}\label{app:D3}

For $\ell=-1-k^2/2$ let $F(t)=1+\ell t^2+ikt$, $g(t)=t-2t^3/3$ and
$f_a=i^jF(t_a)$ on $\B_j$. On $[-T_h,T_h]$, $2T_h^2<1$ and
$\ell^2T_h^2\le2$ give $0\le g'\le1$ and $|F(t)|^2=1-2t^2+\ell^2t^4\le1$.
For a pair of main clusters use $\xi=t_a,\ \zeta=t_b,\ v=\xi-\zeta=d_{ab}$,
$C_{ab}=\xi^2+\xi\zeta+\zeta^2\ge0$ and $G_{ab}=1-2C_{ab}/3$. The trial norms are
\begin{align*}
 |F(\xi)-iF(\zeta)|^2
 &=2-2(\xi^2+\zeta^2)-2kv\\
 &\quad+\ell^2(\xi^4+\zeta^4)+2\ell k\xi\zeta v,\\
 |F(\xi)-F(\zeta)|^2&=k^2v^2+\ell^2(\xi^2-\zeta^2)^2,\\
 |F(\xi)+F(\zeta)|^2&=(2+\ell(\xi^2+\zeta^2))^2+k^2(\xi+\zeta)^2.
\end{align*}
The polynomial checks in \texttt{algebra(k)} verify these and the cancellation identities
\begin{align*}
 g(\xi)-g(\zeta)&=vG_{ab},\\
 1-\xi^2-\zeta^2&=G_{ab}-v^2/3,\\
 C_{ab}v^2&=\xi^4+\zeta^4-\xi^3\zeta-\xi\zeta^3.
\end{align*}
The replay \texttt{extras/appb\_checks.py} in the verification code independently checks these polynomials with SymPy.
Pairing the terms of $\sum_\B 4x_ag(t_a)F_a=0$ gives
\begin{multline*}
 4\sum_{\rm adj}N_{ab}(g_a-g_b)\cos v
 =2\sum_{\rm same}N_{ab}(g_a-g_b)\sin v\\
 -2\sum_{\rm opp}N_{ab}(g_a-g_b)\sin v+\mathrm{Bnd},
\end{multline*}
where $\mathrm{Bnd}=4\sum_{a\in\B,e\in\E}N_{ae}g(t_a)\sin(\varphi_a-\varphi_e)$.
The same and opposite terms have coefficient $2$ on ordered sums, equivalent to $4$ on unordered sums.

Write $S=\sum_{\rm same}x_ax_bv^2$, $A_P=\sum_{\rm adj}N_{ab}v^2$,
$S_j=\sum_{\B_j}x_at_a^2$, $T_j=\sum_{\B_j}x_at_a$ and
$M_4=\sum_\B x_at_a^4\le T_h^2V$. Expansion with $\sum_\B x_at_a=0$ gives
\begin{align}
 \sum_{a<b\in\B}x_ax_bv^4&=(1-b)M_4+3V^2,\nonumber\\
 \sum_{a<b\in\B}x_ax_bC_{ab}v^2&=(1-b)M_4.
 \label{eq:fourth}
\end{align}
In particular $U_4=\sum_{\rm adj}N_{ab}v^4\le(T_h^2+3\Phi)V$.
This bound holds for every unordered subset with weights $N_{ab}$ or $x_ax_b$.
An ordered sum needs twice the bound.

\begin{table}[t]
\caption{Scalar chain in \texttt{check\_final.py}. Monospace entries quote the checker's diagnostic labels, which use an older vocabulary (sheet = group, exceptional = stray).}
\label{tab:shell-chain}
\begin{ruledtabular}
\begin{tabular}{@{}>{\raggedright\arraybackslash}p{0.21\columnwidth}>{\raggedright\arraybackslash}p{0.72\columnwidth}@{}}
Lines & Rational tests used in the proof\\
\hline
111--125
 & Parameter domains, fourth-modulus bound, cutoff sine, sine concentration, main-cluster moment and mean bounds\\
126--140
 & Positive $A_0$, squared coarse-floor bound, \texttt{Opposite cosine lower bound}, \texttt{Bootstrap upper sheet mass}\\
141--144
 & Square-root bound, \texttt{Rebalanced exceptional degree}, \texttt{Ideal exceptional positive-density domain}\\
145--175
 & Positive Young shift, missing-edge domain, leading main-cluster floor, \texttt{Positive sheet denominator}, coupled interval and product domains, \texttt{Coupled Young determinant endpoint}, \texttt{Sheet weight ratio enclosure}\\
176--183
 & Cubic multiplier and trial norm domains, \texttt{Opposite-edge coefficient}, computation of $c_B$\\
186--204
 & Slope $k\ge1$, computation of $K$, positive adjusted main and stray coefficients, absorption determinant\\
\end{tabular}
\end{ruledtabular}
\end{table}

\paragraph*{Remainder.}
In the adjacent contribution, the cancellation
$4(1-\xi^2-\zeta^2)\sin v=4(g(\xi)-g(\zeta))\cos v+R$ has
\begin{multline*}
 R/4=\sin v-v\cos v-v^3/3+(v^2/3)(v-\sin v)\\
 -(2C_{ab}/3)(\sin v-v\cos v).
\end{multline*}
Integrating $1-\cos z$, $z\sin z$ and $z(\sin z-z)$, respectively, gives
\[
 \begin{split}
 |v-\sin v|&\le |v|^3/6,\\
 |\sin v-v\cos v|&\le |v|^3/3,\\
 |\sin v-v\cos v-v^3/3|&\le |v|^5/30.
 \end{split}
\]
Thus $|R|\le(16/45)|v|^5+(8/9)C_{ab}|v|^3$. Since $|v|\le D$,
Eq.~\eqref{eq:fourth} applied to the nonnegative sums gives
\begin{align*}
 \sum_{\rm adj}N_{ab}|R|
 &\le D\{\tfrac{16}{45}[(1-b)M_4+3V^2]+\tfrac89(1-b)M_4\}\\
 &\le D(\tfrac{56}{45}T_h^2+\tfrac{16}{15}\Phi)V.
\end{align*}

\paragraph*{Remaining main-cluster terms.}
The \texttt{Opposite-edge coefficient} check below implies $D<1$ and $4+2\ell D^2>0$ because $-2\ell\ge2$ and $k>0$. The replay \texttt{extras/appb\_checks.py} verifies these two bounds directly.
Since $g$ is increasing and 1-Lipschitz, $0\le(g_a-g_b)\sin v\le v^2$. The virial terms on the right therefore cost at most $2S+\mathrm{Bnd}$, with the opposite term nonpositive.
For the other adjacent terms, the inequalities $v\sin v\ge v^2-v^4/6$ and
$\xi\zeta\ge-v^2/4$ (from $(\xi+\zeta)^2\ge0$) give
\begin{align*}
 -4k\sum_{\rm adj}N_{ab}v\sin v&\le-4kA_P+\tfrac{2k}{3}U_4,\\
 4\ell k\sum_{\rm adj}N_{ab}\xi\zeta v\sin v&\le-\ell kU_4.
\end{align*}
The second bound uses $0\le v\sin v\le v^2$ and $\ell k<0$.
Each cluster sees at most $2M$ mass in its two neighboring groups. Using $\sin v\le D$ gives
\begin{align*}
 2\ell^2\sum_{\rm adj}N_{ab}(\xi^4+\zeta^4)\sin v
 &\le2\ell^2D\sum_{\rm adj}x_ax_b(\xi^4+\zeta^4)\\
 &\le4M\ell^2DM_4\le M\ell^2D^3V.
\end{align*}
The same-group cosine lies in $(0,1]$. The ordered identity
\[
 \sum_{a,b\in\B_j}x_ax_b(t_a^2-t_b^2)^2
 =2X_j\sum_{\B_j}x_at_a^4-2S_j^2
\]
therefore bounds the same-group cost by $k^2S+2M\ell^2T_h^2V$.
For opposite pairs, $\xi^2+\zeta^2\le D^2/2$ and $\ell<0$ give
$|F(\xi)+F(\zeta)|^2\ge4+2\ell D^2>0$. As $\cos v\ge1-D^2/2>0$,
\begin{align*}
 \cos v\,|F(\xi)+F(\zeta)|^2
 &\ge (1-D^2/2)(4+2\ell D^2)\\
 &=4-(2-2\ell)D^2-\ell D^4\\
 &\ge4-(2-2\ell)D^2.
\end{align*}
The opposite-group cosine is $-\cos v$, so its contribution is at most $-[4-(2-2\ell)D^2]O_o$.
Collecting these bounds with the unordered $U_4$ estimate gives the main--main cost
\begin{multline}
 Q_{\B\B}\le(k^2+2)S-4kA_P-[4-(2-2\ell)D^2]O_o\\
 +\mathrm{Bnd}+\mathcal R V,
 \label{eq:QBB}
\end{multline}
where the checker subtracts
\begin{align*}
 \mathcal R&=(2M\ell^2+\tfrac{2k}{3}-\ell k)T_h^2
             +(2k-3\ell k)\Phi+M\ell^2D^3\\
 &\quad+D(\tfrac{56}{45}T_h^2+\tfrac{16}{15}\Phi)
\end{align*}
from $g_B$ (\texttt{check\_final.py}, lines 181--183).

\begin{table}[!tbp]
\caption{Selected exact-check margins, rounded for display. The absorption entries use the rational lower bound $E'$ obtained by replacing $\sqrt2$ with $14143/10000$. The margins for
Eqs.~\eqref{eq:A-S2} and~\eqref{eq:A-S3} are after multiplication by $s_7(2h)^2$.
The coarse-floor margin is squared, the absorption margin is about $1/52$ of $S_*$,
and the near-region box entry is the minimum relative slack of the squared test for Eq.~\eqref{eq:T2}
among QP leaves in the 6486-box cover.}
\label{tab:margins}
\begin{ruledtabular}
\begin{tabular}{ll}
inequality & margin\\
\hline
near/far overlap & the single point $D_4=\Delta$\\
main 2nd moment, Eq.~\eqref{eq:A-S3} & $3.5\times10^{-20}$\\
stray mass, Eq.~\eqref{eq:A-S2} & $9.7\times10^{-15}$\\
coarse group floor (squared) & $3.4\times10^{-14}$\\
coupled Young endpoint & $7.6\times10^{-8}$\\
absorption, $4B'E'-S_*$ & $2.35\times10^{-8}$\\
opposite-edge coefficient & $1.8\times10^{-3}$ (out of 4)\\
near-region boxes (relative slack) & $1.2\times10^{-7}$\\
\end{tabular}
\end{ruledtabular}
\end{table}

\subsection{Missing edges and coupled Young weights}\label{app:D4}

Set $A_1=\sum_{\rm adj}x_ax_bv^2$. For $a\in\B_j$ its missing adjacent mass
$\delta_a=\sum_{b\text{ adjacent to }a}x_b(1-P_{ab})$ satisfies
\[
 \delta_a\le1-\mus-X_{j+2}+o_a\le D_j+o_a.
\]
The first bound subtracts missing opposite mass from total missing mass, and the second uses $M\ge1-\mus$ and $D_j=M-X_{j+2}\ge0$.
For positive $w_j$ and adjacent groups $j,i$, Young's inequality gives
$v^2\le(1+w_i/w_j)\xi^2+(1+w_j/w_i)\zeta^2$.
Thus the contribution at cluster $a$ to the upper bound on $A_1-A_P$ is
at most $(1+\max_{i\sim j}w_i/w_j)(D_j+o_a)x_at_a^2$.
Expansion and centering give
\begin{align*}
 S&=2\sum_j(X_jS_j-T_j^2)\le2\sum_jX_jS_j,\\
 A_1&=\sum_j(1-b-X_j-X_{j+2})S_j-2\sum_jT_jT_{j+1}\\
 &\ge\sum_j(1-b-X_j-X_{j+2})S_j,
\end{align*}
since $-2\sum_jT_jT_{j+1}=2(T_0+T_2)^2$.
Writing $-4kA_P=-4kA_1+4k(A_1-A_P)$ and using $X_{j+2}+D_j=M$, the $S_j$ coefficient is at most
\[
 \mathcal A X_j+C+4kD_j\max_{i\sim j}(w_i/w_j),
\]
where $\mathcal A=2k^2+4+4k$ and $C=4k(M-1+\beta)$. With $w_i/w_j\le R_w$,
the $o_a$ cost is at most $4k(1+R_w)T_h^2O_o$ since $\sum_\B x_ao_a=O_o$.

Put $L=-g_B-C$, so $L-\mathcal AM>0$ by the scalar test.
For adjacent $j,i$ we need
\begin{equation}
 (L-\mathcal AX_j)(L-\mathcal AX_i)\ge16k^2(D_j+\tau)(D_i+\tau).
 \label{eq:C2}
\end{equation}
For $s=X_j+X_i$, the caps give $|X_j-X_i|\le2M-s$. Applying Eq.~\eqref{eq:S2} in both orientations gives $|X_j-X_i|\le3s-2c'$ and $s\ge2c'/3$. The complementary adjacent pair gives $s\le1-2c'/3$.
By AM--GM,
\begin{align*}
 (D_j+\tau)(D_i+\tau)&\le\bigl(M+\tau-(1-b-s)/2\bigr)^2\\
 &\le\bigl(M+\tau-(1-\beta-s)/2\bigr)^2,
\end{align*}
since the first bracket is $(D_j+D_i)/2+\tau>0$ and increases with $b$.
The left side of Eq.~\eqref{eq:C2} equals
$(L-\mathcal As/2)^2-\mathcal A^2(X_j-X_i)^2/4$.
Then Eq.~\eqref{eq:C2} follows from $F_*(s)\ge0$, where
\begin{multline*}
 F_*(s)=(L-\mathcal As/2)^2
 -\tfrac{\mathcal A^2}{4}\min(2M-s,3s-2c')^2\\
 -16k^2\bigl(M+\tau-(1-\beta-s)/2\bigr)^2.
\end{multline*}
The branches meet at $s_*=(M+c')/2$ and have leading coefficients $-2\mathcal A^2-4k^2$ and $-4k^2$, respectively. Their concavity reduces the test to $2c'/3,s_*,1-2c'/3$, the three points of the checker's \texttt{Coupled Young determinant endpoint} test.

Take $w_j=\sqrt{(D_j+\tau)/(L-\mathcal AX_j)}$. Equation~\eqref{eq:C2} gives
\begin{align*}
 D_j\frac{w_i}{w_j}
 &\le\sqrt{\frac{(D_j+\tau)(D_i+\tau)(L-\mathcal AX_j)}{L-\mathcal AX_i}}
 \le\frac{L-\mathcal AX_j}{4k}.
\end{align*}
This makes the $S_j$ coefficient at most $L+C=-g_B$.
Since $m_3\le X_j\le M$ and $0\le D_j\le M-m_3$, all ratios are bounded by the checked $R_w$ with
\[
 R_w^2\ge
 \frac{M-m_3+\tau}{\tau}\frac{L-\mathcal Am_3}{L-\mathcal AM}.
\]
The total opposite coefficient is nonpositive by
$4-(2-2\ell+k(1+R_w))D^2\ge0$. Dropping it and using $\sum_jS_j=V$ in Eq.~\eqref{eq:QBB} yields
\[
 Q_{\B\B}\le-c_BV+\mathrm{Bnd},\qquad c_B=g_B-\mathcal R.
\]

\subsection{Stray clusters and the box checks}\label{app:D5}

For fixed $e\in\E$, measure its angle from $\bar u$, then rotate by quarter turns and reflect if needed to fold it to $p\in[h-\alpha,\pi/4]$. The angular distance to the nearest corner is 1-Lipschitz, so its original distance $|u_e|>h$ differs from $p$ by at most $|\bar u|$. In the folded frame the main offsets are $t_a$ or $-t_a$, with the same bounds, centering and second moment. Use those offsets and relabeled groups in this subsection. Set $s=\sin p$, $c=\cos p$, and $W_j=\sum_{a\in\B_j}x_aP_{ea}$. Degree and equilibrium give
\begin{gather*}
 \sum_j W_j\ge\mus-b,\qquad 0\le W_j\le X_j\le M,\\
 |s(W_2-W_0)+c(W_1-W_3)|\le U_t+b.
\end{gather*}
The torque error uses $\sum_\B x_aP_{ea}|t_a|\le U_t$ for replacing main phases by their corners, and $b$ for stray links.

\paragraph*{Torque correction and ideal cost.}
Set $y=(s/c)(W_0-W_2)$, which lies in $[-M,M]$ because $s/c\le1$ and $W_j\in[0,M]$, and $\delta=y-(W_1-W_3)$.
To increase the difference, increase entry 1 and then decrease entry 3, with capacity $M-(W_1-W_3)$.
To decrease it, decrease entry 1 and then increase entry 3, with capacity $M+(W_1-W_3)$.
Since $|y|\le M$, the correction stays in $[0,M]^4$, balances torque exactly and costs
\[
 \|\widetilde W-W\|_1=|\delta|\le(U_t+b)/c.
\]
Using $c\ge1/\sqrt2$, the exact degree tests give
\[
 \sum_j\widetilde W_j\ge\mus-\beta-\sqrt2(\alpha+\beta)\ge q>2M.
\]
Put $A=\widetilde W_0+\widetilde W_2$, $B=\widetilde W_1+\widetilde W_3$,
$H_e=Ac+iBs$, and $z_e=H_e/|H_e|=\chi+i\psi$. Then $A,B\ge q-2M>0$ and $\chi,\psi\ge0$.
With
\[
 r=|(\widetilde W_0-\widetilde W_2)+i(\widetilde W_1-\widetilde W_3)|\le\sqrt2M,
\]
define $G_{\rm ideal}(W,z)=\sum_j W_j\cos(j\pi/2-p)|z-i^j|^2$.
Expanding the square gives
\begin{multline*}
 \tfrac12G_{\rm ideal}(\widetilde W,z_e)
 =c(\widetilde W_0-\widetilde W_2)+s(\widetilde W_1-\widetilde W_3)\\
 -\operatorname{Re}(\bar z_e H_e)\le r-|H_e|.
\end{multline*}
Undo the fold on $z_e$ to define $f_e$ in the global frame.

\paragraph*{Leaf bounds and ideal penalty.}
The root rectangle is $[h-\alpha,3927/5000]\times[0,(14143/10000)M]$ in $(p,r)$.
The checker uses Machin-type bounds for $3927/5000>\pi/4$ and checks $\sqrt2<14143/10000$.
It rebuilds the complete binary cover from the paths, bisecting the angle if its width exceeds four times the radial width, and the radius otherwise.
It rejects overlapping paths and missing leaves. See the README of the verification code.
Choose the leaf $[p_L,p_U]\times[r_L,r_U]$ containing $(p,r)$, breaking boundary ties by lexicographic path order. Set
\begin{align*}
 c_L&=1-p_U^2/2+p_U^4/24-p_U^6/720,\\
 s_L&=s_7(p_L),\qquad s_U=p_U-p_U^3/6+p_U^5/120,\\
 A_L&=\max(r_Lc_L,q-2M+r_Ls_L),\\
 B_U&=2M-r_Ls_L,\qquad R_L=\max(1,B_U/A_L),\\
 z_L&=\min(1,R_Ls_U/c_L),\\
 C_L&=4(1+kR_L)+4(1+z_L)/c_L,\\
 \sigma_L&=s_7(2(p_L-\alpha)),\qquad W_L=C_L/\sigma_L.
\end{align*}
The alternating series and monotonicity give $0<c_L\le c$ and $0<s_L\le s\le s_U$.
Torque balance gives $|\widetilde W_0-\widetilde W_2|=rc$ and
$|\widetilde W_1-\widetilde W_3|=rs$, so
\[
 rc\le A\le2M-rc,\qquad rs\le B\le2M-rs.
\]
Because $A+B\ge q$, we have $A\ge A_L>0$, $B\le B_U$, hence $B/A\le R_L$.
In particular $c\psi=(B/A)s\chi\le R_Ls$ and $\psi\le\min(1,R_Ls/c)\le z_L$.

Adding the lower and upper bounds for $A,B$ shows $r(c+s)\le2M<q\le A+B$.
Thus $A,B$ can be decreased within those intervals to $A',B'$ with $A'+B'=q$.
Then $x=A'$ lies in $[\mathrm{lo},\mathrm{hi}]$, where
\begin{align*}
 \mathrm{lo}&=A_L,&
 \mathrm{hi}&=\min(2M-r_Lc_L,q-r_Ls_L).
\end{align*}
A leaf marked empty has $\mathrm{lo}>\mathrm{hi}$ and is therefore infeasible.
On a QP leaf the rationals $a_0,g_0,\lambda_{\rm lo},\lambda_{\rm hi}$ satisfy
$\lambda_{\rm lo},\lambda_{\rm hi}\ge0$ and the polynomial identity
\begin{multline}
 c_L^2x^2+s_L^2(q-x)^2-g_0=(c_L^2+s_L^2)(x-a_0)^2\\
 +\lambda_{\rm lo}(x-\mathrm{lo})+\lambda_{\rm hi}(\mathrm{hi}-x).
 \label{eq:shell-QP}
\end{multline}
The right side is nonnegative on the interval. Since $A'\le A$ and $B'\le B$,
\[
 |H_e|^2\ge c_L^2x^2+s_L^2(q-x)^2\ge g_0
 \ge\bigl(r_U+\tfrac\gamma2 W_L^2s_U^2\bigr)^2.
\]
The checker recomputes the leaf quantities. It tests the interval, multiplier signs and positive sine charge, then checks \texttt{QP linear coefficient},
\texttt{QP constant coefficient} and \texttt{Derivative-weighted ideal penalty}.
With $r\le r_U$, $s\le s_U$ and the ideal cost bound, this proves
\begin{equation}
 \begin{split}
 |H_e|-r&\ge\tfrac\gamma2 W_L^2s_U^2,\\
 G_{\rm ideal}(\widetilde W,z_e)&\le-\gamma W_L^2s^2.
 \end{split}
 \label{eq:T2}
\end{equation}

\paragraph*{Taylor bounds and edge accounting.}
For a main link in group $j$ put $\mathcal D_j(t)=|z_e-i^jF(t)|^2$ and
\[
 L_j(t)=2\cos(j\pi/2+t-p)\mathcal D_j(t)+4g(t)\sin(j\pi/2+t-p).
\]
The first term counts both orientations and the second includes this link's share of $\mathrm{Bnd}$.
Under unfolding, rotation multiplies both trial entries by the same unit factor. Reflection conjugates both entries since $F(-t)=\overline{F(t)}$, and changes the signs of $g$ and the sine together. Thus the folded expression gives the actual cost.

Differentiating at zero gives
\begin{align*}
 L_0'(0)=L_2'(0)&=-4(s\chi+kc\psi),\\
 L_1'(0)=L_3'(0)&=4(c\psi+ks\chi).
\end{align*}
Using $0\le\chi\le1$ and $c\psi\le R_Ls$, the absolute bounds are $4(1+kR_L)s$ for the even groups and $4(R_L+k)s$ for the odd groups. The latter is at most the former since $(k-1)(R_L-1)\ge0$. The absolute bounds allow signed offsets.

At zero the weighted sum is $\sum_\B x_aP_{ea}L_{j(a)}(0)=2G_{\rm ideal}(W,z_e)$.
Only entries 1 and 3 change in the torque correction, with $L_1(0)=4s(1-\psi)$ and $L_3(0)=-4s(1+\psi)$.
Its error is at most
\[
 4s(1+\psi)\|W-\widetilde W\|_1
 \le \frac{4s(1+z_L)}{c_L}U_t+8\sqrt2\,sb,
\]
where the $b$ term uses $1+\psi\le2$ and $c\ge1/\sqrt2$.
Adding the absolute derivative cost gives $C_Ls\sqrt V+8\sqrt2\,sb$.
Now $|u_e|\ge p_L-\alpha>0$ since $p_L\ge h-\alpha$ and $h>2\alpha$.
The alternating sine bound at $2(p_L-\alpha)\le\pi/2$ gives
$0<\sigma_L\le\sin(2(p_L-\alpha))\le\sin(2|u_e|)$.
Thus the first-order and correction errors cost at most
$W_Ls\sin(2|u_e|)\sqrt V+8\sqrt2\,sb$.

From $\mathcal D_j(t)=2-2t^2+\ell^2t^4-2\operatorname{Re}(\bar z_e i^jF(t))$,
differentiation and $\ell<0$ give, for $|t|\le T_h$,
\begin{align*}
 |\mathcal D_j'|&\le d_1=4T_h+4\ell^2T_h^3+2k-4\ell T_h,\\
 |\mathcal D_j''|&\le d_2=4+12\ell^2T_h^2-4\ell.
\end{align*}
Using $\mathcal D_j\le4$, $|g|\le T_h$, $|g'|\le1$ and $|g''|\le4T_h$, twice differentiating $L_j$ gives
$|L_j''|\le16+20T_h+4d_1+2d_2=2K$.
This $K$ is the expression in \texttt{check\_final.py}, lines 191--193.
The Taylor remainder is at most $Kt_a^2$ per link. With $W_e=W_L$ and $s_e=\sin p_e$, Eq.~\eqref{eq:T2} and the Taylor bounds give
\begin{multline}
 \text{Contribution of }e\le x_e\bigl[-2\gamma W_e^2s_e^2
 +W_es_e\sin(2|u_e|)\sqrt V\\
 +8\sqrt2\,s_eb+KV\bigr].
 \label{eq:econtrib}
\end{multline}
This includes both orientations of each main link and its boundary term exactly once.
The stray--stray ordered edges cost at most $4b^2$ since every $|f_e|=1$.

\subsection{Absorption}\label{app:absorb}

Let $J=\sum_\E x_eW_e^2s_e^2$ and $C_{\min}=8+4k$.
From $R_L\ge1$, $z_L\ge0$ and $c_L\le1$, we have $C_L\ge C_{\min}$.
Since $\sigma_L\le\sin(2(p_L-\alpha))\le\min(1,2\sin p_e)$, it follows that
$W_e\ge C_{\min}$, $W_es_e\ge C_{\min}/2$, and $W_e^2s_e\ge C_{\min}^2/2$.
Thus $J\ge C_{\min}^2b/4$ and $\sum_\E x_es_e\le2J/C_{\min}^2$.
By Cauchy--Schwarz and Eq.~\eqref{eq:S1}, the cross terms in Eq.~\eqref{eq:econtrib} total at most $\sqrt{S_*VJ}$.
The remaining errors $8\sqrt2\,b\sum_\E x_es_e$, $4b^2$ and $KbV$ are bounded by
$16\sqrt2\,\beta J/C_{\min}^2$, $16\beta J/C_{\min}^2$ and $K\beta V$, respectively.
Adding the main-cluster bound gives
\[
 Q(f)\le-B_VV-E_JJ+\sqrt{S_*VJ},
\]
where $B_V=c_B-K\beta$ and $E_J=2\gamma-16(\sqrt2+1)\beta/C_{\min}^2$.
For $d=10^{-10}$ let $B'=B_V-16d$ and $E'=E_J-16d/C_{\min}^2$.
The checker tests $B'>0$, $E'>0$, $4B'E'\ge S_*$ with the smaller rational stray coefficient obtained by replacing $\sqrt2$ with $14143/10000$.
By AM--GM, $B'V+E'J\ge2\sqrt{B'E'VJ}\ge\sqrt{S_*VJ}$.
Using $4b\le16J/C_{\min}^2$ gives $Q(f)\le-d(16V+4b)$.
These inequalities also apply when $\E$ is empty and $b=J=0$, and use no division by $V$ or $J$.
Finally,
\[
 D_4\le2(1-\operatorname{Re}(e^{-4i\bar u}\rho_4))\le16V+4b.
\]
The first inequality uses $|1-e^{-4i\bar u}\rho_4|^2\ge0$. The second uses
$1-\cos4t\le8t^2$ on $\B$ and $1-\cos v\le2$ on $\E$.
This proves Eq.~\eqref{eq:shellQ}. \qed

\section{Reproduction}\label{app:repro}\label{app:formats}

From the extracted \texttt{release\_0717/} directory, \texttt{./run\_checks.sh} runs \texttt{check\_final.py}
on \texttt{shell\_717.json}, runs \texttt{far/verify4.py} on the $310$-band cover (each far leaf names its
band, threshold and cutoffs; Appendix~\ref{app:far} states the acceptance test), checks the $323$ hashes in
\texttt{SHA256SUMS}, and writes \texttt{verify4\_out.json}. The Python checkers use only the standard
library. The feasible moment vector of Sec.~\ref{sec:limits} is in \texttt{primal\_peak.json}, with its
checks recorded in \texttt{EXACT\_PEAK.json}; to replay it and the supplementary identities, run
\texttt{python3 -B appb\_checks.py} from the adjacent \texttt{extras/} directory (SymPy required).

\end{document}